\documentclass[twocolumn,floatfix]{revtex4-2}%
\usepackage{graphicx}%
\usepackage{amsmath,amssymb}%
\usepackage{amsfonts}
\usepackage{amssymb}
\usepackage{mathrsfs}
\usepackage{color}
\usepackage{bm}
\usepackage{ulem}

\def\s{{\sigma}}
\def\e{{\epsilon}}
\def\k{{ {\bm k} }}
\def\p{{ {\bm p} }}
\def\q{{ {\bm q} }}
\def\Q{{ {\bm Q} }}

\def\0{{ {\bm 0} }}
\def\w{{\omega}}
\def\a{{\alpha}}

\allowdisplaybreaks[4]

\begin{document}
\title{
Paramagnon-Interference Mechanism for Three-Dimensional Bond Order in Kagome Metals AV$_3$Sb$_5$
(A=Cs, Rb, K): Analysis by the Density-Wave Equation
}
\author{
Seiichiro Onari$^{1}$, Rina Tazai$^{2}$, Youichi Yamakawa$^{1}$,
and Hiroshi Kontani$^1$
}
\date{\today }   

\begin{abstract}
The mechanism of CDW and its 3D structure are important fundamental issues in kagome metals.
We have previously shown that, based on a 2D model, $2\times 2$ bond order (BO) emerges due to the paramagnon-interference (PMI) mechanism and that its fluctuations lead to $s$-wave superconductivity. 
This paper studies these issues based on realistic 3D models of kagome metals AV$_3$Sb$_5$ (A=Cs, Rb, K).
We reveal that a commensurate 3D $2\times 2\times 2$ BO is caused by the PMI mechanism, by performing the 3D density-wave (DW) equation analysis for all A=Cs, Rb, K models in detail. 
Our results indicate a BO transition temperature $T_{\rm BO}\sim 100$K within the regime of moderate electron correlation. 
The 3D structure of BO is attributed to the three-dimensionality of the Fermi surface, while the 3D structure of BO is sensitively changed, since the Fermi surface is quasi-2D.
Based on the analysis of the DW equation, by taking into account a finite third-order Ginzburg-Landau (GL) term, (i) shift stacking $2\times 2\times 2$ BO can be realized via a first-order transition below $T_{\mathrm{BO}}$. Here, the in-plane BO pattern  (tri-hexagonal or star-of-David) is determined by the sign of the third-order GL term, with hole doping tending to favor the tri-hexagonal state. 
On the other hand, if the third-order GL term is very small, (ii) alternating vertical stacking BO may instead be realized via a second-order transition.
The present study enhances our understanding of the rich variety of BOs observed experimentally.
It is confirmed that the PMI mechanism is the essential origin of the 3D CDW of kagome metals.
\end{abstract}

\address{
$^1$Department of Physics, Nagoya University,
Nagoya 464-8602, Japan \\
$^2$Yukawa Institute for Theoretical Physics, Kyoto University,
Kyoto 606-8502, Japan
}
\sloppy

\maketitle

\section{Introduction}
The multistage rich quantum phase transitions in the kagome-lattice superconductors AV$_3$Sb$_5$ (A=Cs, Rb, K) attract significant attention in the field of condensed matter physics \cite{kagome-exp1,kagome-exp2,kagome-P-Tc1}.
In each V-Sb plane, the 2$\times$2 charge density wave (CDW) has been observed below $T_{\rm BO}\approx 90$ K at ambient pressure
\cite{STM1,STM2}.
The structure of CDW is tri-hexagonal (TrH) or star-of-David (SoD) caused by the triple-$\q$ ($3Q$) bond order (BO), which is the even-parity modulation in the hopping integral $\delta t_{i,j}^{\rm b}$ (=real)
\cite{Thomale2013,SMFRG,phonon-kagome,Thomale2021,Neupert2021,Balents2021,Nandkishore,Tazai-kagome,Tazai-kagome2}.
Below $T_{\rm BO}$, superconductivity (SC) appears at $T_{\rm SC} \approx 1 \sim 3$ K,
and it increases to 8 K under $P\approx3 {\rm GPa}$ \cite{kagome-P-Tc1}.
At ambient pressure for A=Cs,
the highly anisotropic $s$-wave SC state is realized
\cite{Roppongi,SC2},
which is naturally understood based on the BO fluctuation mechanism \cite{Tazai-kagome}.

\begin{figure}[htb]
\includegraphics[width=.99\linewidth]{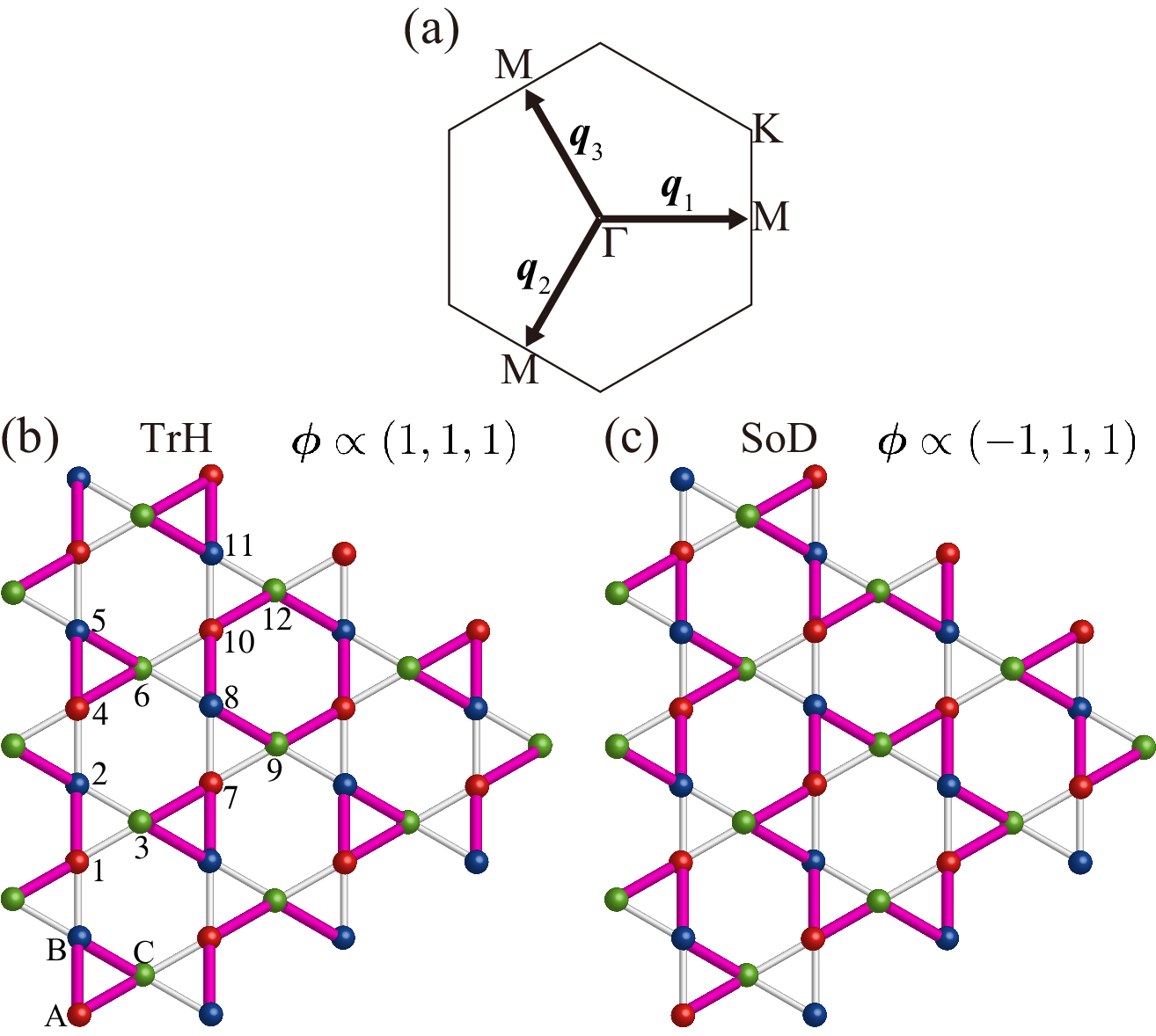}
\caption{
(a) 2D wavevector $\bm{q}_m$ $(m=1,2,3)$ in the original 2D
 Brillouin zone.
(b) 2D TrH BO that corresponds to ${\bm\phi}\propto (1,1,1)$.
In the BO state, the thick purple bond $(i,j)$ corresponds to a decrease in V-V bond length.
(c) 2D SoD BO that corresponds to ${\bm\phi}\propto (-1,1,1)$.
}
\label{fig:fig1}
\end{figure}

In a single V-Sb plane,
the order parameters of the $3Q$ BO are expressed as
${\bm \phi}=(\phi_1,\phi_2,\phi_3)$,
where $\phi_m$ is the strength of the $m$th BO at wavevector $\q_m$ ($m=1,2,3$)
as shown in Fig. \ref{fig:fig1} (a).
The TrH and SoD BOs are shown in Figs. \ref{fig:fig1} (b) and (c), respectively.
Each $3Q$ order is composed of 4 unit cells of the kagome lattice, that is, 12 V atoms.

It is important to understand the BO state 
because it is an essential electronic property in kagome metals.
Many theoretical researches have been performed.
The BO is formed in the presence of strong off-site interaction $I$.
Large $I$ is mediated by the effective off-site electron correlations
\cite{Thomale2013,SMFRG,Neupert2021,Balents2021,Tazai-kagome}.
The bond-stretching phonon also mediates $I$
\cite{phonon-kagome}.
The absence of acoustic phonon anomaly at $T_{\rm BO}$
\cite{Kohn}
indicates the important contribution from the electron correlations.
In Ref. \cite{Tazai-kagome}, the authors found that the 
BO is naturally induced by the ``paramagnon-interference (PMI) mechanism'', 
which is dropped in the mean-field-type approximations.
This mechanism naturally explains the orbital and bond orders 
observed in Fe-based and cuprate superconductors 
\cite{Kontani-AdvPhys}.

\begin{figure}[htb]
\includegraphics[width=.99\linewidth]{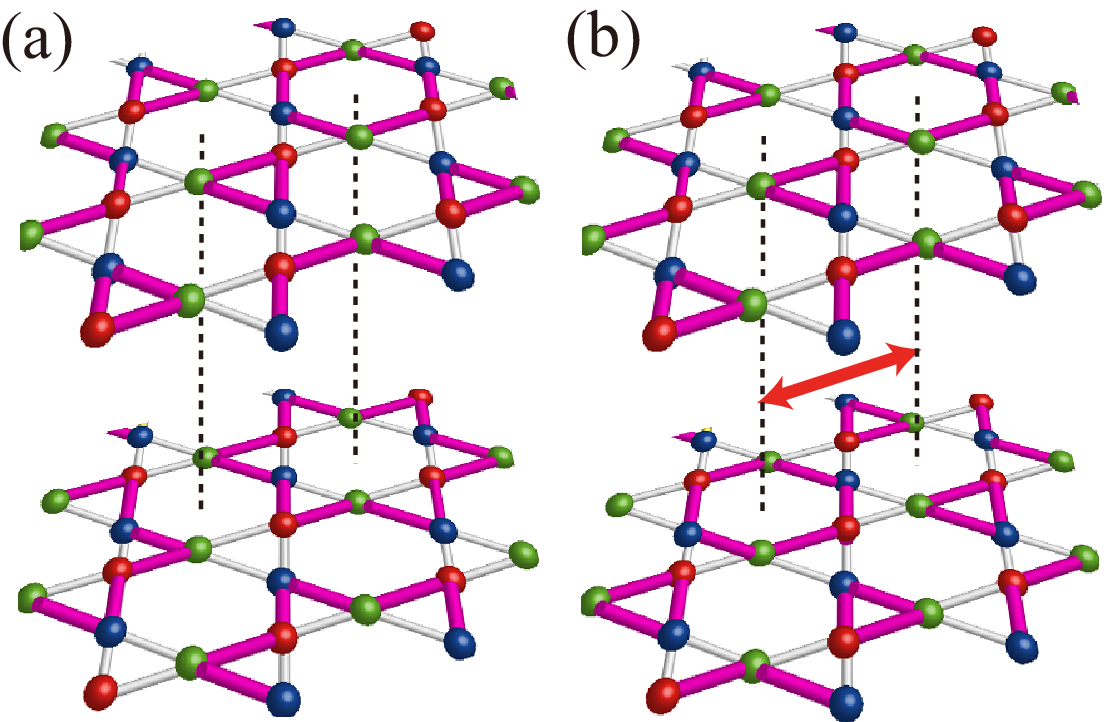}
\caption{
Possible $2\times2\times2$ 3D BO states:
(a) ``alternating v-BO'' with $C_6$ symmetry.
(b) ``TrH s-BO'' with nematic $C_2$ symmetry.
The red arrow represents the nematic director.
}
\label{fig:fig2}
\end{figure}

Recently, the three-dimensional (3D) nature of the BO in kagome metals has attracted increasing attention because it drastically alters their electronic properties.
Depending on the carrier doping and the pressure, the BO states with period-1, period-2, and period-4 along the $c$-axis,
which are called the $2\times2\times C$ BO ($C=1,2,4$),
have been reported.
As shown in Figs. \ref{fig:fig2} (a) and (b), in the case of $C=2$,
TrH-SoD alternating vertical stacking BO (v-BO) is realized for
$\{q_1^z,q_2^z,q_3^z\}=\{\pi,\pi,\pi\}$,
while the nematic TrH or SoD shift stacking BO (s-BO) is realized for $\{q_1^z,q_2^z,q_3^z\}=\{\pi,\pi,0\}$, where $q_m^z$ is
the $z$ component of wavevector in the $m$th BO.
In the nematic s-BO, $C_6$ symmetry is broken to $C_2$. 
The nematic director is represented by the arrow connecting the $C_6$
center of the upper layer to that of the lower layer, as shown in Fig. \ref{fig:fig2} (b).

In addition,
the time-reversal-symmetry-breaking TRSB state has been observed by various experimental methods \cite{STM1,muSR2-K,muSR4-Cs,muSR5-Rb,AHE1,AHE2,eMChA,Moll-hz,Asaba-torque}.
As for the origin of the TRSB state, the chiral charge current order, which is the pure imaginary BO
$\delta t_{i,j}^{\rm c}$ (=imaginary),
has been proposed \cite{Neupert2021,Balents2021,Nandkishore,Tazai-kagome2,Fernandes-CLC,Nat-g-ology}. 
This topic attracts significant attention in kagome metals.
The recent theoretical study \cite{Tazai-kagome2} has revealed that the current order structure is strongly 
coupled to the background BO.
Thus, the 3D nature of the BO is very important for understanding nematicity and the chiral current order.

In this paper, we study the 3D nature of the density wave (DW) in kagome metals based on the realistic 3D multiorbital models for AV$_3$Sb$_5$ (A=Cs, Rb, K), by using the 3D DW equation method
\cite{Tazai-kagome,Onari-FeSe,Onari-AFN,Onari-Ni}.
We reveal that commensurate $2\times 2\times 2$ BO is realized by the PMI mechanism described by the Aslamazov-Larkin vertex corrections (AL-VCs) in all A=Cs, Rb, K models. 
The 3D structure of BO originates from that of the Fermi surface.
 Based on the analysis of the DW equation, by taking into account a finite third-order Ginzburg-Landau (GL) term, (i) $2\times 2\times 2$ s-BO can be realized via a first-order transition below
 $T_{\mathrm{BO}}$. Here, the in-plane BO pattern  (TrH or  SoD) is determined by the sign of the third-order GL term, with hole doping tending to favor the TrH state. 
On the other hand, if the third-order GL term is very small, (ii) alternating v-BO may instead be realized via a second-order transition.
These results are robust for all A=Cs, Rb, and K models.
The present study enhances our understanding of the rich variety of BOs observed experimentally.
It is confirmed that the PMI mechanism is the essential origin of the 3D CDW of kagome metals.

Here, we present a brief summary of the CDW
structure observed in AV$_3$Sb$_5$.
In each layer, the $2\times2$ superlattice structure 
that indicates the formation of the $3Q$ CDW
has been observed for A=Cs, Rb, and K
\cite{STM1,exp13,exp15,exp16,exp17,exp18,exp19,exp20,exp21}.
Recently, the $2\times2\times1$ superlattice
\cite{exp12},
the $2\times2\times2$ superlattice
\cite{Kohn,exp12,exp1,exp2,exp3,exp4,exp5,exp6,exp8,Raman,exp9,XRD,exp22,exp23},
and the $2\times2\times4$ superlattices 
\cite{exp12,XRD,exp22,exp23,exp10,exp11}
have been observed in the CDW phase. In this study, we take into account the $2\times 2\times 1$ and $2\times 2\times 2$ CDW states for simplicity.

In the $2\times2\times2$ CDW state, (i) the nematic TrH or SoD s-BO has been supported by the STM \cite{exp5}, the NMR \cite{exp2,exp3,exp4,exp6,exp18,NMR-SoD}, the ARPES \cite{exp1,exp8}, the Raman \cite{Raman}, and the X-ray \cite{Kohn,exp9,XRD} measurements for A=Cs, Rb, and K.
In general, it is difficult to distinguish between the TrH and SoD BOs
in experiments such as STM and X-ray measurements, since the observed results can be interpreted by both the TrH and SoD BOs.
 On the other hand, (ii) the alternating v-BO has been proposed by combining the ARPES measurement and the DFT first-principles calculation in A=Cs \cite{exp8}.
Consensus on the BO states in kagome metals has yet to be reached, since the observed BO depends on doping, pressure, and experimental method.

Theoretically, the TrH stacking CDW due to the electron-phonon interaction mechanism is obtained by the DFT first-principles studies
\cite{phonon-kagome,theo2,theo3,theo4}.
However, the change in bond length obtained by the first-principles studies is rather large compared to the experimental results \cite{exp12,XRD,exp10}.
Thus, it is reasonable to assume that the origin of the CDW is the electron correlation.
As for the electron correlation mechanism, the 3D CDW formation due to the on-site and off-site Coulomb interaction has been studied based on the mean-field theory
\cite{theo6,theo7}
and the variational method \cite{theo9},
by introducing sizable off-site interactions.

In contrast, the present authors studied the realistic 2D kagome metal model
based on the DW equation method by considering only the on-site Coulomb
interaction $U$ \cite{Tazai-kagome}. In the PMI mechanism, the effective
off-site interaction is induced by the AL terms, which are the higher-order electron correlation.
We revealed that the commensurate $2\times2$ CDW 
is naturally and robustly given by the PMI mechanism. In addition, $s$-wave superconductivity is
explained by the CDW fluctuations.
In this paper, we show that the same mechanism can explain the 3D CDW based on realistic 3D kagome metal models.

GL theories in 2D kagome metals have revealed that the $3Q$
$2\times 2$ TrH or SoD CDW state is stabilized by the 3rd-order GL terms
\cite{Balents2021,Tazai-kagome2,theo7}. The GL theories also expect the coexistence of the 3Q CDW and chiral-charge-current orders. In the coexisting
state, $C_2$ nematicity appears \cite{Tazai-kagome2,theo7}.

The 3D structure of CDW has been discussed in
Refs. \cite{Balents2021,theo5,theo8} by using the 3D GL theory.
The $2\times 2\times 2$ s-BO has been proposed by the 3D GL theory.
In CsV$_3$Sb$_5$, the TrH
s-BO was proposed \cite{theo5,theo8}, where the coefficients of the GL terms are estimated by the first-principles calculation.

\section{3D Multiorbital Model for Kagome Metal}

Here, we derive a first-principles realistic tight-binding model 
for CsV$_3$Sb$_5$, by using the WIEN2k \cite{WIEN2k} and Wannier90
\cite{Wannier90} software. Hereafter, the unit of energy is eV unless otherwise noted.
The derived kinetic term is
\begin{eqnarray}
H_0=\sum_{\k,lm\s}h_{lm}^0(\k) c_{\k l\s}^\dagger c_{\k m\s},
\label{eqn:H0}
\end{eqnarray}
where $\k$ is the 3D wavevector,
$l,m$ represents the V $d$-orbitals and Sb $p$-orbitals, and $\s$ is the
spin index. This is a 30 orbital tight-binding model.
To improve the tight-binding model, we introduce the energy shift of all
Sb $p$-orbitals
$\Delta E_p=-0.2$. The band structure compared to that for $\Delta E_p=0$ is
shown in Fig. \ref{fig:figS2} (a) in Appendix A.
Energy at the $\Gamma$ point $E_\Gamma \sim -0.4$ given for original $\Delta E_p=0$
is reduced to $E_\Gamma \sim -0.55$ by introducing $\Delta E_p=-0.2$.
 Since $E_\Gamma \sim -0.55$ well reproduces the ARPES
measurements \cite{exp8,ARPES-VHS}, $\Delta E_p=-0.2$ is
reasonable for the model of CsV$_3$Sb$_5$.
In this paper, based on the first-principles calculation, the thick purple bond in Figs
\ref{fig:fig1} and \ref{fig:fig2}, where the V-V bond length decreases,
corresponds to a modulation of the hopping integral
$\delta t>0$. 

\begin{figure}[htb]
\includegraphics[width=.99\linewidth]{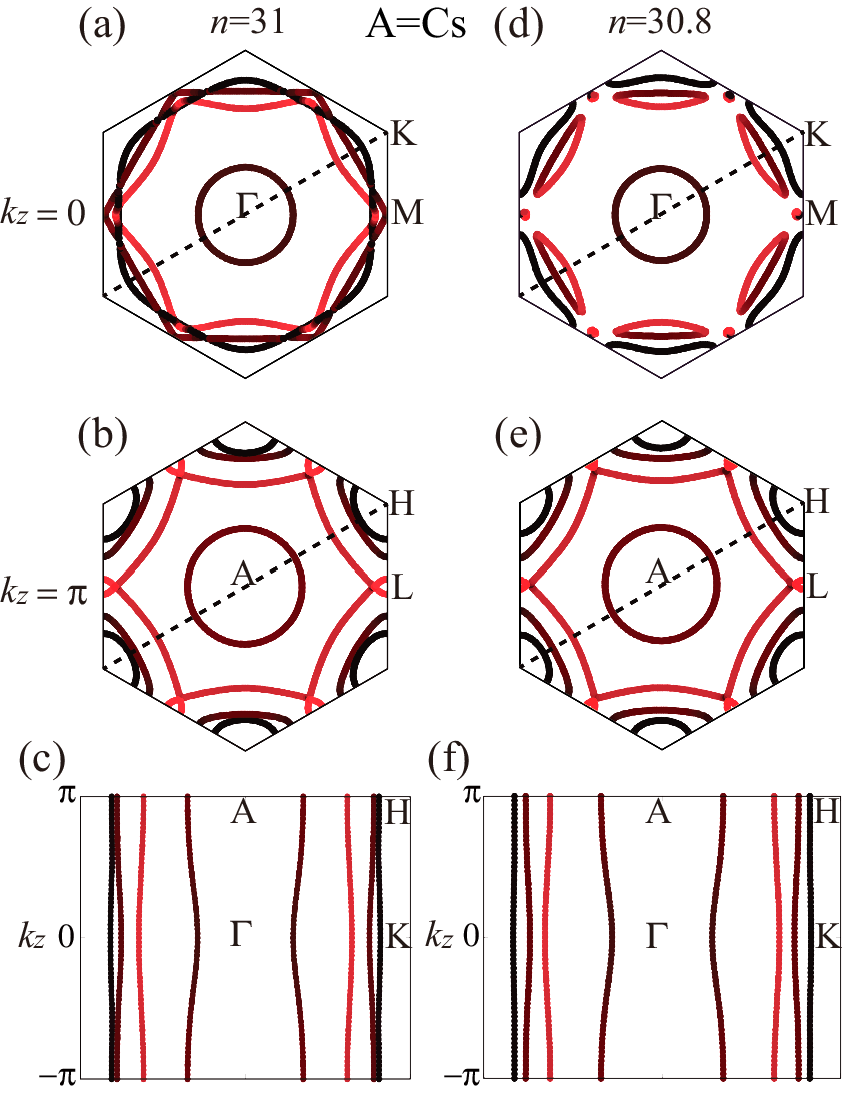}
\caption{
FS structure of the present kagome metal model for A=Cs.
The FS at $n=31$ on the (a) $k_z=0$ plane,
(b) $k_z=\pi$ plane, and (c) plane including $\Gamma$KHA points along the
 dotted line in (a) and (b). The red color denotes the $b_{3g}$-orbital
 weight.
The FS at $n=30.8$ on the (d) $k_z=0$ plane,
(e) $k_z=\pi$ plane, and (f) plane including $\Gamma$KHA points.
}
\label{fig:fig3}
\end{figure}

\begin{figure}[htb]
\includegraphics[width=.99\linewidth]{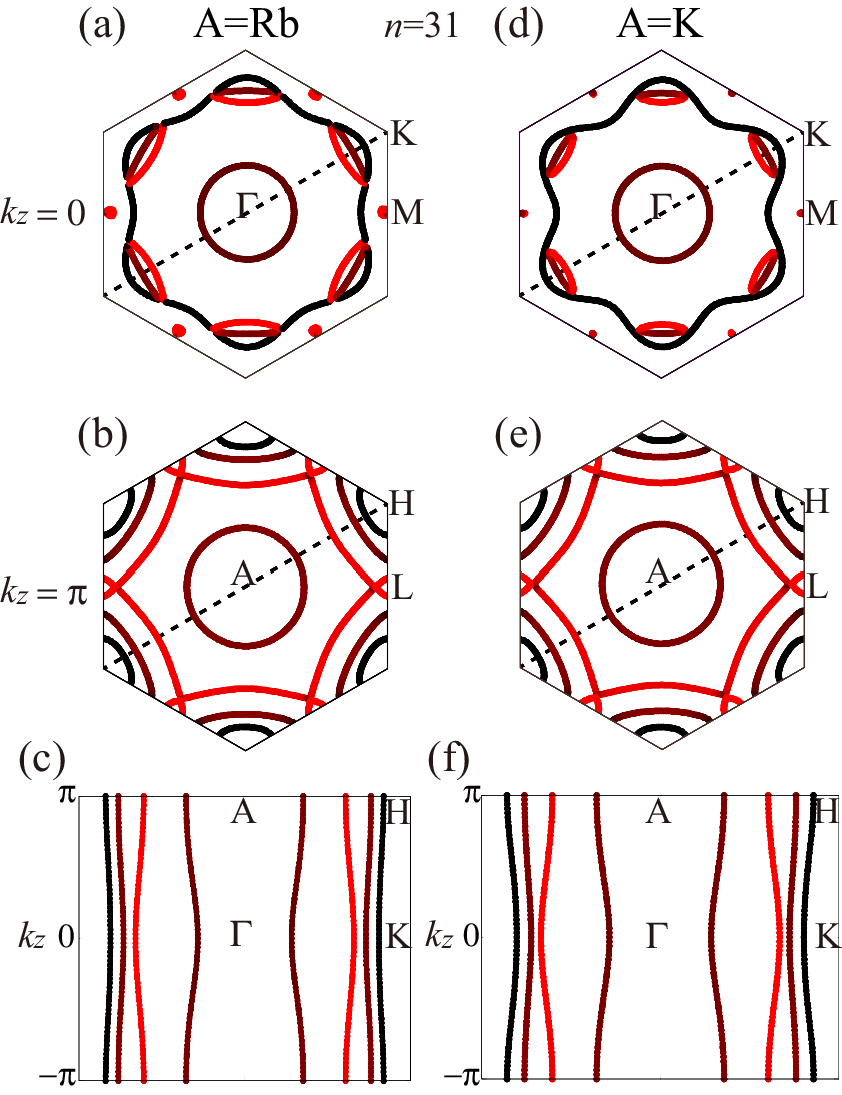}
\caption{
FS structure of the present kagome metal models for A=Rb, K at $n=31$.
The FS for A=Rb on the (a) $k_z=0$ plane,
(b) $k_z=\pi$ plane, and (c) plane including $\Gamma$KHA points. The red color denotes the $b_{3g}$-orbital weight. 
The FS for A=K on the (d) $k_z=0$ plane,
(e) $k_z=\pi$ plane, and (f) plane including $\Gamma$KHA points.
}
\label{fig:figS9}
\end{figure}

The 3D FS structure is shown in Fig. \ref{fig:figS2} (b) in Appendix A.
The FSs at $n=31$ (=original electron number) are 
shown in Figs. \ref{fig:fig3} (a)-(c).
Also, the FSs at $n=30.8$ (hole doping) are 
shown in Figs. \ref{fig:fig3} (d)-(f). These FSs are quasi-2D.
The FSs in $k_z=0$ are different between $n=31$ and $n=30.8$ due to the
Lifshitz transition around the M point, while the FSs in $k_z=\pi$ at
$n=31$ are
almost the same as those at $n=30.8$.
The FSs of A=Rb, K for $E_p=-0.2$ are shown in Fig. \ref{fig:figS9}.
The FSs of A=Rb, K are similar to those of the hole-doped A=Cs in
Fig. \ref{fig:fig3}.
The band structures of A=Rb, K for $E_p=-0.2$ are shown in
Fig. \ref{fig:figS8} in Appendix A. They are similar
to that of A=Cs. $E_\Gamma \sim -0.6$ for A=Rb, K is also consistent
with the ARPES measurements \cite{exp8}.

Next, we calculate the effects of the strong Coulomb interaction of $3d$-orbital electrons on V sites.
In kagome metals, the FSs are mainly composed of
two of the five $d$-orbitals of V atom:
The ``pure-type'' FS is composed of three
$d_{xz}$ (or $b_{3g}$ rep.) orbitals on sublattices A, B, C,
and the ``mix-type'' FS is composed of three
$d_{yz}$ orbitals (or $b_{2g}$ rep.).
In the pure-type FS, the band structure around each van Hove singularity
vHS-$S$ ($S$=A, B, C) is composed of the single sublattice $S$,
which is known as the ``sublattice interference''.
In contrast, in the mix-type FS, each vHS-$S$ is composed of 
two of three sublattices.
Both the pure-type and mix-type bands are nearly independent, although they weakly hybridize each other.

Here, we study the spin-channel and charge-channel 
susceptibilities, ${\hat\chi}^s(q)$ and ${\hat\chi}^c(q)$,
based on the random-phase-approximation (RPA),
which is equivalent to the mean-field approximation.
To perform the diagrammatic calculation, we introduce the $30\times30$ matrix expression of the Green function as
\begin{eqnarray}
{\hat G}(k)=[(i\e_n+\mu){\hat 1}-{\hat h}^0(\k)]^{-1},
\label{eqn:G0}
\end{eqnarray}
where $\mu$ is the chemical potential, $k\equiv (\k,\e_n)$, and $\e_n=(2n+1)\pi T$ is the fermion Matsubara frequency.
In the RPA, ${\hat\chi}^s(q)$ and ${\hat\chi}^c(q)$ are given as
\begin{eqnarray}
{\hat\chi}^{s(c)}(Q)&=&{\hat\chi}^{0}(q)\left[{\hat 1}-{\hat\Gamma}^{s(c)}{\hat\chi}^{0}(Q)\right]^{-1},\\
\chi^{0}_{ll',mm'}(Q)&=&-\frac{T}{N}\sum_{k}G_{lm}(k+Q)G_{m'l'}(k),
\end{eqnarray}
where $Q\equiv(\Q,\w_l)$ with the bosonic Matsubara frequency $\w_l=2\pi T l$,
and $\chi^{0}_{ll',mm'}(Q)$ is the irreducible susceptibility.
Here, $\Gamma_{ll'mm'}^{s(c)}$ is the 5-orbital Coulomb interaction 
for the same V sublattice in the spin (charge) channel, 
composed of the intra-orbital $U$, inter-orbital $U'$, 
and the exchange term $J=(U-U')/2$.
The expression of $\Gamma_{ll'mm'}^{s(c)}$ is presented in
Ref. \cite{Kontani-PRL2010,Onari-PRL2012}. We use $\k$ meshes
$N=N_x\times N_y\times N_z=60\times60\times32$, and Matsubara frequency
$N_\w=512$. 
Here, we introduce the spin (charge) Stoner factor $\a_{S(C)}$
given as the largest eigenvalue of ${\hat\Gamma}^{s(c)}{\hat\chi}^{0}(q)$.
$\chi^{s(c)}$ diverges when $\a_{S(C)}$ reaches unity.
In the RPA, $\a_{S}$ is always larger than $\a_C$ \cite{Kontani-PRL2010,Onari-PRL2012} for $U>U'$, so the nonmagnetic BO in kagome metals cannot be explained
unless large nearest-neighbor Coulomb interaction $V$ ($V>0.5U$) exists.
However, the BO and the current order are naturally explained 
by considering the beyond-mean-field electron correlations described by the VCs.

In previous theoretical studies \cite{Tazai-kagome},
both the spin and BO fluctuations develop in pure-type and mix-type FSs. 
It was shown that the electron correlations are stronger on pure-type FS 
composed of $d_{xz}$ (or $b_{3g}$) orbitals.
Therefore, in the present study,
we consider the on-site Coulomb interactions only on the $b_{3g}$ orbitals.
Here, we assign $l=1,2,3$ to the $b_{3g}$ orbital on sites A, B, C, respectively.
In this case, the Coulomb interaction is simply given as
${\hat \Gamma}^{s(c)}_{mm',ll'}=U^{s(c)}\delta_{m,m'}\delta_{l,l'}\delta_{l,m}$,
where $U^s=U$ and $U^c=-U$.
Then, the RPA susceptibility is simplified as
\begin{eqnarray}
{\hat{\tilde\chi}}^{s(c)}(Q)&=&{\hat{\tilde\chi}}^{0}(Q)\left[{\hat 1}-(+)U{\hat{\tilde\chi}}^{0}(Q)\right]^{-1},
\end{eqnarray}
where ${\hat{\tilde\chi}}^{s(c)}(q)$ is $3\times3$ matrix:
${\tilde\chi}_{l,m}^0(Q)=\chi^{0}_{ll,mm}(Q)$ with $l,m=1,2,3$.
In the next section, we introduce the DW equation to study the BO in kagome metals.
The kernel function of the DW equation 
is composed of the RPA susceptibility.

Hereafter, we denote a 3D variable wavevector $\Q=(\q,q^z)$,
where $\q$ ($q^z$) is the 2D (1D) variable wavevector.
We also introduce the notation $\Q_m=(\q_m,q_m^z)$ ($m=1,2,3$),
where $\q_m$ is the fixed 2D wavevector introduced in Fig. \ref{fig:fig1} (a),
while $q_m^z$ is treated as a variable.

We introduce the ``form factor'' of the BO at wavevector $\Q$, 
${\hat g}_{\Q}(k)$, which is given by the Fourier transformation 
of the 3D BO, such as the 3D stacking of Figs. \ref{fig:fig1} (b) and
(c). 
The form factor is normalized by setting ${\rm
max}|{\hat g}_{\Q}(k)|=1$.
Here, ${\hat g}_{\Q}(k)$ is a $3\times3$ matrix 
with respect to the three sublattices A, B, and C. 
In this study, we derive the BO form factor and its eigenvalue $\lambda_\Q$
in 3D momentum space microscopically, 
by analyzing the DW equation for the 3D kagome metal models.
It is found that both ${\hat g}_{\Q_m}(k)$ and $\lambda_{\Q_m}$ ($m=1,2,3$)
have very small $q_m^z$-dependences.

\section{DW Equation Analysis for 3D Kagome Metal Model}

\begin{figure}[htb]
\includegraphics[width=.99\linewidth]{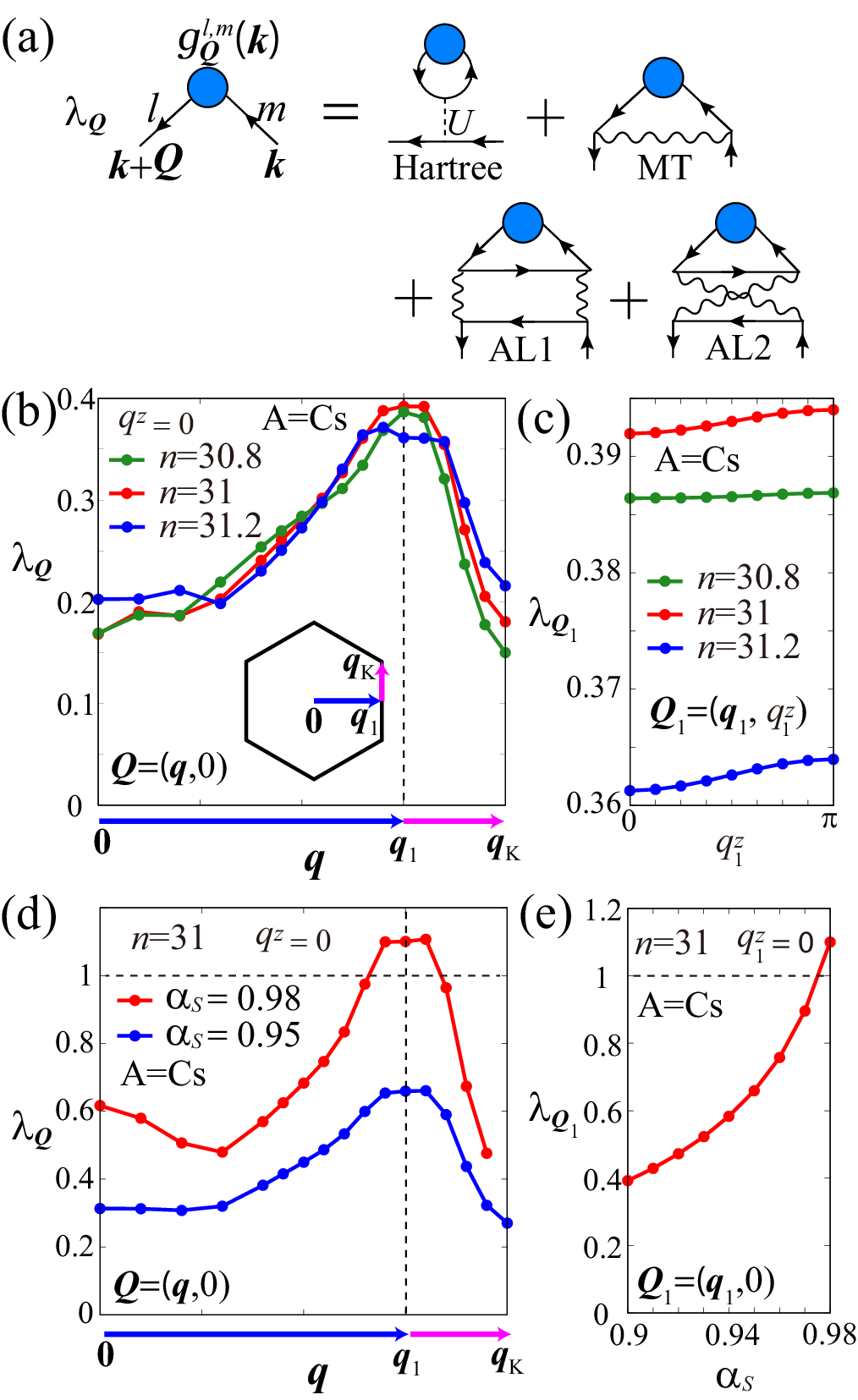}
\caption{
(a) Charge-channel density-wave (DW) eigenvalue equation.
The largest eigenvalue $\lambda_\Q$ represents the instability of 
the charge-channel DW at wavevector $\Q$,
and the form factor $f$ gives the order parameter.
The wavy lines represent the spin susceptibilities.
Two AL terms describe the paramagnon interference mechanism 
that gives various unconventional DW states.
 (b-c) Obtained $\lambda_\Q$ for A=Cs at $n=30.8,31,31.2$:
 (b) $\lambda_\Q$ in the $q^z=0$ plane, where $\q_K=\left(\frac{2\pi}{\sqrt{3}},\frac{2\pi}{3}\right)$.
 (c) Small $q_1^z$-dependence of $\lambda_{\Q_1}$ for $\Q_1=(\q_1,q_1^z)$.
 The BO solution is obtained for any $q_1^z$.
 (d) $\lambda_{\Q}$ for $\alpha_S=0.95,0.98$ in $q^z=0$ at $n=31$ of A=Cs.
 (e) $\alpha_S$ dependence of $\lambda_{\Q_1}$ for $q_1^z=0$ at $n=31$
 of A=Cs.
}
\label{fig:fig4}
\end{figure}

Here, we study the BO in the 3D kagome metal models
based on the DW equation theory. 
In this theory, various nonmagnetic DW orders are caused
by the beyond-RPA non-local correlations. 
This theory was first developed to explain the orbital nematic 
state in Fe-based superconductors, and it has been successfully applied to 
the cuprates \cite{Yamakawa-Cu} and the twisted-bilayer graphene \cite{Onari-TBG}. 
Recently, the present authors studied 2D kagome metals
in Ref. \cite{Tazai-kagome}.
By solving the DW equation, including the Hartree, Maki-Thompson (MT), and
AL terms shown in Fig. \ref{fig:fig4} (a),
we can calculate the infinite series of AL and MT terms.
The AL terms represent the PMI mechanism,
which gives rise to the BO in kagome metals.
The analytic expression of the charge-channel DW equation is
\begin{eqnarray}
\lambda_{\Q}g_\Q^{L}(k)&=& -\frac{T}{N}\sum_{p,M_1,M_2}
I_\Q^{L,M_1}(k,p) 
\nonumber \\
& &\times \{ G(p)G(p+\Q) \}^{M_1,M_2} g_\Q^{M_2}(p) ,
\label{eqn:DWeq}
\end{eqnarray}
where $k\equiv (\k,\e_n)$ and $p\equiv (\p,\e_m)$.
The largest eigenvalue $\lambda_\Q$ represents the DW instability 
at wavevector $\Q$, and the corresponding form factor ${\hat g}_\Q$
gives the DW order function.
$L\equiv (l,l')$ and $M_i$ represent the pair of sublattice indices $A,B,C$.
The phase factor of the eigenvector is fixed to satisfy
the Hermitian condition $g_\Q^{ll'}(\k,\e_n)=[g_{-\Q}^{l'l}(\k+\Q,-\e_n)]^*$.
In addition, we fix the sign of ${\hat g}_{\Q_m}$
so that $\phi_m {\hat g}_{\Q_m}$
reproduces the BO shown in Fig. \ref{fig:fig1}.

The effective interaction ${\hat I}_\Q(k,p)$ in the DW equation is given as
\begin{eqnarray}
&& I^{l l', m m'}_{\Q} (k,k')
= \sum_{b = s, c} \frac{a^b}{2} 
\Bigl[ -V^{b}_{l m, l' m'} (k-k') 
\nonumber \\
&&
+\frac{T}{N} \sum_{p} \sum_{l_1 l_2, m_1 m_2}
	V^{b}_{l l_1, m m_1} \left( p+\Q \right) 
	V^{b}_{m' m_2, l' l_2} \left( p \right)
 \nonumber \\
&& \qquad\qquad\qquad \quad
\times G_{l_1 l_2} (k-p) G_{m_2 m_1} (k'-p)
 \nonumber \\
&& 
+\frac{T}{N} \sum_{p} \sum_{l_1 l_2, m_1 m_2}
	V^{b}_{l l_1, m_2 m'} \left( p+\Q \right) 
	V^{b}_{m_1 m, l' l_2} \left( p \right)
 \nonumber \\
&& \qquad\qquad\qquad 
\times G_{l_1 l_2} (k-p) G_{m_2 m_1} (k'+p+\Q) \Bigr] ,
\label{eqs:kernel} 
\end{eqnarray}
where $a^{s(c)} = 3$($1$) and $p = (\p,\w_l)$. 
$\hat{V}^{b}$ is the $b$-channel interaction given by 
$\hat{V}^{b} = \hat{\Gamma}^{b} + \hat{\Gamma}^{b} \hat{\chi}^{b} \hat{\Gamma}^{b}$. 

Note that the form factor is proportional to the 
particle-hole condensation 
$\sum_\s \langle c_{\k+\q,l,\s}^\dagger c_{\k,l',\s}\rangle$,
or equivalently, the symmetry-breaking component in the self-energy. 

Here, we explain the obtained numerical results for A=Cs
at $T=0.04$ and $U=2.6$, where $\a_S=0.90$.
Figure \ref{fig:fig4} (b) shows the obtained $\lambda_\Q$ on the $q^z=0$ plane
at $n=30.8,31,31.2$:
$\lambda_\Q$ shows the almost commensurate peak at $\q= \q_1$,
which corresponds to the 2D BO wavevector in kagome metals.
Interestingly, the $q_1^z$-dependence of $\lambda_{\Q_1}$ is quite small
($\lesssim 1$\%), as shown in Fig. \ref{fig:fig4} (c),
reflecting the quasi-2D character of the FSs in kagome metals.
We note that the value of $\lambda_{\Q_1}$ for
$q_1^z=\pi$ is slightly larger than that for $q_1^z=0$ since the spin
susceptibility peaks in $q^z\sim\pi$ due to the 3D structure of FSs, which derives the wavevector with
$q^z\sim\pi$ as discussed in Ref. \cite{Onari-AFN}. In this case, the
alternating v-BO with $\{q_1^z,q_2^z,q_3^z\}=\{\pi,\pi,\pi\}$ is the
most favorable, while the TrH (SoD) s-BO with
$\{q_1^z,q_2^z,q_3^z\}=\{\pi,\pi,0\}$ is also favored. Within the DW
equation, we cannot distinguish between the TrH s-BO and the SoD s-BO.
{The band structures at $n=30.8$ and $31.2$ are shown in Figs. \ref{fig:fig15}(a) and (b) in Appendix A, respectively. The energy at the  $b_{3g}$-orbital vHS $(E_{\rm vHS})$ is obtained as $E_{\rm vHS}=-0.013(0.079)$ at $n=30.8(31.2)$. At $n=31.2$, $T<|E_{\rm vHS}-\mu|$ is satisfied. Even in the condition, a nearly $2\times 2\times 2$ BO is reproduced.}
Figures \ref{fig:fig4} (d) and (e) show $\alpha_S$
dependence of $\lambda_{\Q}$ for
$q^z=0$ at $n=31$, $T=0.04$. $\lambda_{\Q_1}$ reaches $1$ for $\alpha_S \gtrsim
0.97$. Thus, the BO transition temperature
 $T_{\rm BO}\sim 100$K is obtained in the moderate electron correlation.

\begin{figure}[htb]
\includegraphics[width=.99\linewidth]{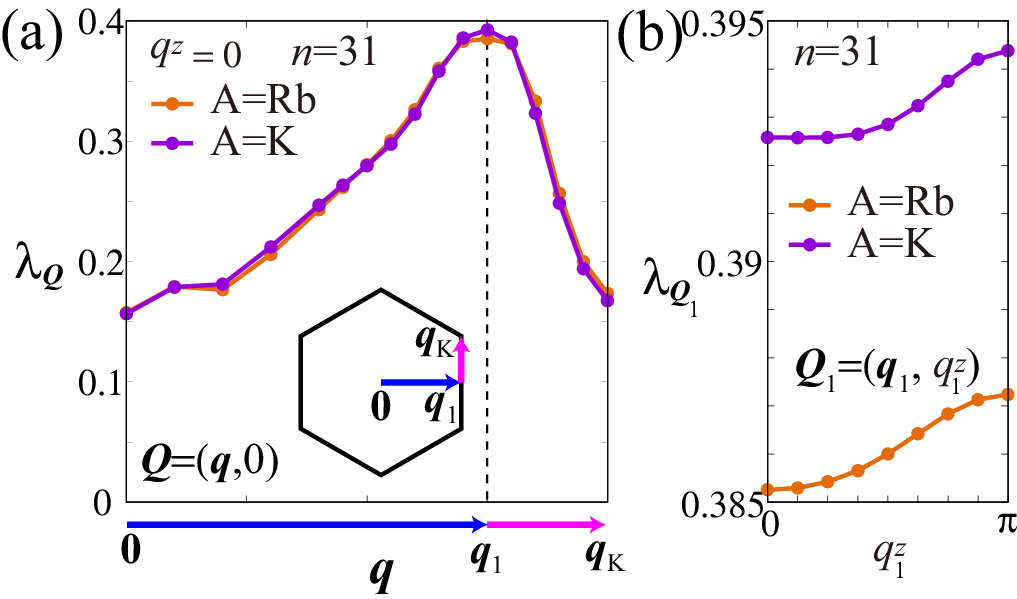}
\caption{
(a-b) Obtained $\lambda_\Q$ for A=Rb, K at $n=31$:
 (a) $\lambda_\Q$ in the $q^z=0$ plane.
 (b) Small $q_1^z$-dependence of $\lambda_{\Q_1}$ for $\Q_1=(\q_1,q_1^z)$.
}
\label{fig:figS6}
\end{figure}

In deriving Figs. \ref{fig:fig4} (b)-(e),
we take only the Coulomb interaction on the $b_{3g}$ orbital into account.
This simplification does not alter the qualitative results,
while the quantitative results are improved by considering
the multiorbital Coulomb interaction.

{It is noteworthy that the total BO eigenvalue is enhanced as $\lambda_{\bm{q}}^{\rm tot}=\lambda_{\bm{q}}+\gamma_{\bm{q}}$, where $\gamma_{\bm{q}} (>0)$ originates from the electron–phonon interaction, as discussed for Fe-based superconductors \cite{Fe-phonon}. Therefore, even when $\lambda_{\bm{q}}$ is slightly incommensurate (for $n=31.2$), $\lambda_{\bm{q}}^{\rm tot}$ can become commensurate due to the $\bm{q}$-dependence of $\gamma_{\bm{q}}$.
Furthermore, the $3Q$ BO transition in kagome metals is a weak first order due to a symmetry-allowed third-order term in the GL free energy. As a result, even if $\lambda_{\bm{q}}$ is slightly incommensurate, a commensurate $2\times 2\times 2$ BO may emerge through a first-order transition driven by lattice deformation.
It is expected that the $2\times 2\times 2$ BO in kagome metals arises from the cooperative effects of Coulomb repulsion and electron–phonon interactions.} 

Figures \ref{fig:figS6} (a) and (b) show $\Q$ dependence of
$\lambda_\Q$ for A=Rb, K at $T=0.04$ with $\a_S=0.90$ for $U=2.5$. The obtained results are similar to those for
A=Cs shown in Fig. \ref{fig:fig4}. Thus, it is robust that the alternating v-BO and the TrH (SoD) s-BO are favored in kagome metals.

\begin{figure}[htb]
\includegraphics[width=.9\linewidth]{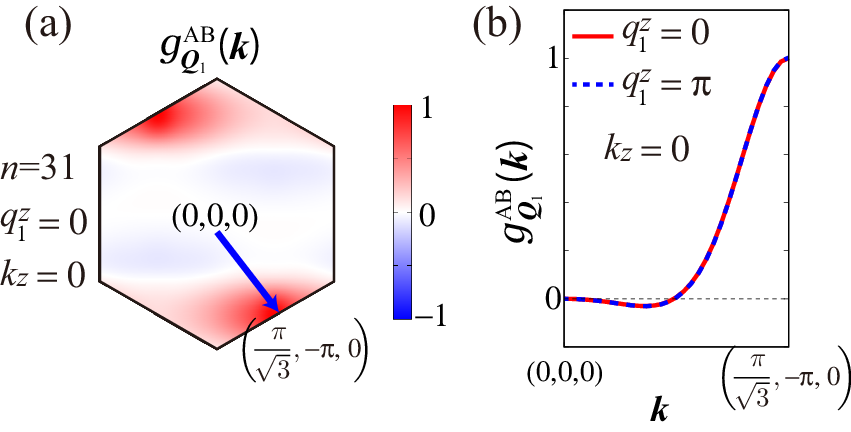}
\caption{
(a) Obtained form factor $g_{\Q_1}^{\rm AB}(\k)$ 
 in the $k_x$-$k_y$ plane at $\q_1^z=0$ at $n=31$.
(b) $\k$ dependence of $g_{\Q_1}^{\rm AB}(\k)$ for $q_1^z=0,\pi$ along the blue arrow in (a).
}
\label{fig:fig5cd}
\end{figure}

Figure \ref{fig:fig5cd} (a) shows the 
obtained form factor $g_{\Q_1}^{\rm AB}(\k)$ 
in the $k_x$-$k_y$ plane at $\q_1^z=0$ 
in the case of $n=31$. Here, $\hat{g}_{\Q}(\k)\equiv
\hat{g}_{\Q}(\k,0)$ is given by the analytic continuation (linear extrapolation) of $\hat{g}_{\Q}(k)$.
We also show the 
$\k$ dependence of $g_{\Q_1}^{\rm AB}(\k)$ for $q_1^z=0,\pi$ at $k_z=0$
in Fig. \ref{fig:fig5cd} (b). $q_1^z$ dependences of
$g_{\Q_1}^{\rm AB}(\k)$ are quite small. We confirm that $k_z$
dependences of $g_{\Q_1}^{\rm AB}(\k)$ is also negligible. Thus,
we ignore the $q^z$ and $k_z$ dependences of the form factors.

To summarize, the eigenvalue for the BO $\lambda_\Q$ 
derived from the DW equation analysis
for the 3D multiorbital kagome metal models for A=Cs, Rb, K
shows the maximum value at commensurate $\Q=\Q_m$ ($m=1,2,3$),
while its $q_m^z$ dependence is small.
This result would be consistent with the
rich variety of the 3D BO states observed in kagome metals.
In the present study, $\lambda_{(\q_m,\pi)}$ is slightly
larger than $\lambda_{(\q_m,0)}$.
In this case, the
alternating v-BO with $\{q_1^z,q_2^z,q_3^z\}=\{\pi,\pi,\pi\}$ in
Fig. \ref{fig:fig2} (a) is the
most favorable, while the TrH (SoD) s-BO with
$\{q_1^z,q_2^z,q_3^z\}=\{\pi,\pi,0\}$ in Fig. \ref{fig:fig2} (b) is also
favored.
These results are robust for all A=Cs, Rb, and K models.
The rich variety of BOs observed in
 experiments would be understood by the competition between the
alternating v-BO and TrH (SoD) s-BO. We verify that the PMI mechanism in the DW equation is the essential origin of the 3D CDW in
 kagome metals.

\section{Discussion}
Here, we discuss the 3D structure of CDW in kagome metal based on the
obtained results by the 3D DW equation and the 3rd-order GL free energy.
For simplicity, we take account of commensurate $q_m^z=-q_m^z(=0,\pi)$
BO wavevectors, which correspond to $2\times 2\times 1$ and $2\times 2\times 2$ BO.
From the $\Q$ dependence of $\lambda_\Q$ shown in Figs. \ref{fig:fig4}
and \ref{fig:figS6},
we find that $\lambda_{\q_1,\pi}$ is slightly larger than
$\lambda_{\q_1,0}$.  For this reason, the commensurate $2\times 2\times 2$ alternating
v-BO with $\{q_1^z,q_2^z,q_3^z\}=\{\pi,\pi,\pi\}$ and nematic TrH (SoD)
BO with $\{q_1^z,q_2^z,q_3^z\}=\{\pi,\pi,0\}$ are favored based on the 3D DW equation due to the PMI mechanism.
On the other hand, the coefficient of the 3rd-order GL term $b_1(q_1^z,q_2^z,q_3^z)$ is shown in Fig. \ref{fig:fig6} (b) in Appendix B.
The obtained negative (positive) $b_1$ favors the TrH (SoD) BO.
We confirmed that $b_1(\pi,\pi,0)$ is almost the same as $b_1(0,0,0)$
since $q^z$ dependence of $b_1$ is very small.
Thus, $2\times 2\times 2$ TrH (SoD) s-BO competes with $2\times 2\times
1$ TrH (SoD) BO.
Combined with the results of the 3D DW equation, the nematic
$2\times 2\times 2$ TrH (SoD) s-BO is realized via a first-order transition below
 $T_{\mathrm{BO}}$ by the 3rd-order GL free
energy. In contrast, the contribution from the 3rd-order GL free
energy is absent to the alternating
v-BO with $\{q_1^z,q_2^z,q_3^z\}=\{\pi,\pi,\pi\}$, which violates the constraint $q_1^z+q_2^z+q_3^z=0$ (modulo $2\pi$).
The alternating
v-BO will be realized via a second-order transition if the 3rd-order GL free energy is very small.
Based on the $n$ dependence of $b_1$, the TrH BO tends to be stabilized
in the hole-doped regime, whereas the SoD BO tends to be favored in the electron-doped regime. {We summarize these results for A=Cs in Table \ref{Table1}.}
However, the obtained value of $b_1$ is sensitive to $T$. 
Quantitative evaluation of $b_1$ is an important future issue.

\begin{table}[h]
\footnotesize
\begin{tabular}{|c|c|c|c|}\hline
A=Cs & Hevily h-doped & Nearly undoped & Heavily e-doped \\ \hline
3rd-order GL & $b_1<0$ & $b_1\sim 0$ & $b_1>0$ \\ \hline
DW+GL & TrH s-BO & alt. v-BO & SoD s-BO \\ \hline
\end{tabular}
\caption{Summary of the results for A=Cs. By combining the results of (i) DW Eq. $\lambda_{(\bm{q}_1,\pi)}\gtrsim \lambda_{(\bm{q}_1,0)}$ and (ii) 3rd-order GL term $b_1$, the TrH s-BO, the alternating v-BO, and the SoD s-BO are favored in the heavily hole-doped, nearly undoped, and heavily electron-doped cases, respectively.}
\label{Table1}
\end{table}
\normalsize

In the NMR study for A=Cs \cite{exp18,NMR-SoD}, the SoD BO has been proposed at low
temperatures, while other NMR studies have proposed the TrH s-BO near the CDW transition temperature.
Since the value of $b_1$ at $n=31$ is not robust, as shown in Figs. \ref{fig:fig6} (b), the TrH and SoD BOs can compete.
These BOs are favored by the DW equation due
to the PMI mechanism, where the
alternating v-BO is also favored. These competitions would explain the rich variety of the 3D DW states in kagome metals.

Finally, we clarify the relation between a modulation of the V-V length
$\delta a$
and a BO parameter $\delta t$. We compare the band structure obtained from
a first-principles calculation using $\delta a$ with that obtained from
a model calculation using $\delta t$.  In the following, we set $\Delta E_p=0$. The relation between
 $\delta a$ and $\delta t$ is essentially unaffected by the value of
 $\Delta E_p$.
 
\begin{figure}[htb]
\includegraphics[width=.99\linewidth]{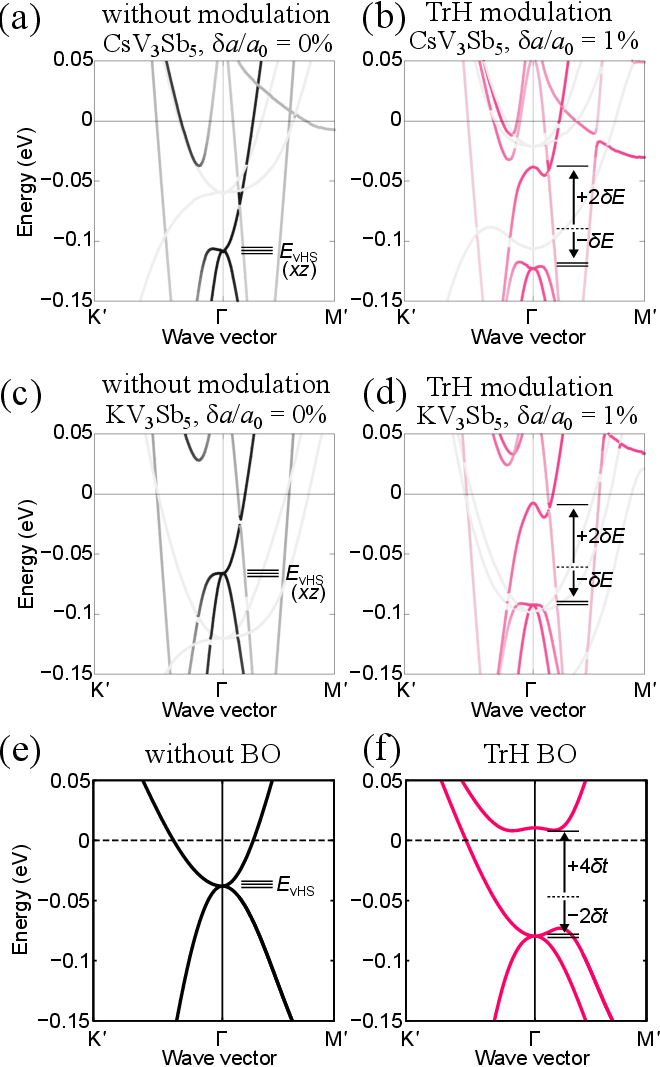}
 \caption{
2D folded $2\times 2$ band structures of CsV$_3$Sb$_5$ obtained by the first-principles
 calculation: (a) without the V-V length modulation $\delta a=0$, and (b) with TrH-type modulation $\delta
 a/a_0=1$\%.
 The weight of the $d_{xz}$ orbital is represented by the color intensity.
 Panels (c) and (d) present analogous results for KV$_3$Sb$_5$,
 corresponding to those shown for CsV$_3$Sb$_5$ in panels (a) and (b).
 (e) 2D folded $2\times 2$ band structure given by the simple model without BO, and (f) that with TrH BO. $\delta t>0$ denotes the
 BO parameter on the thick purple bond for the TrH BO.
}
\label{fig:fig13}
\end{figure}

Figures \ref{fig:fig13} (a) and (b) present the 2D folded $2\times 2$ band
structures for A=Cs obtained from the first-principles
calculations using $\delta a=0$ and the TrH-type modulation $\delta
a/a_0=0.01$, respectively. $a_0$ denotes the
 original V-V length without modulation. Hereafter, we focus on the $d_{xz}$ orbital,
the weight of which is represented by color intensity in these figures.
The $2\times 2$
TrH-type modulation shown in Fig. \ref{fig:fig1} (b) is
characterized by a shortened bond length $a_0-\delta a$ on the thick
purple bond and an elongated bond length $a_0+\delta a$ on other bonds.
 Due to the $2\times 2$ band folding, three vHSs at M points in the
 original Brillouin zone are folded to $\Gamma$ point.
In the $2\times 2$ TrH-type modulation, the triply degenerate bands at $\Gamma$ point split into one upper
and two lower bands. 
This band splitting is also observed for A = K, as shown in Figs. \ref{fig:fig13}
(c) and (d). The main difference between the results for A=Cs and A=K is the energy of vHS $E_{\rm vHS}$;
$E_{\rm vHS}\sim -0.05$eV for A=K is higher than $E_{\rm vHS}\sim -0.1$eV for A=Cs.

Here, we compare the band splitting obtained by the first-principles
calculation using the TrH-type modulation with that obtained by
a model calculation using the BO parameter $\delta t$.
The band splittings by the TrH-type modulation in Figs. \ref{fig:fig13}
(b) and (d) are effectively reproduced using a
simple model with a positive $\delta t>0$, as shown in
Figs. \ref{fig:fig13} (e) and (f).
In the simple model introduced in
Ref. \cite{Tazai-kagome}, only the three $d_{xz}$ orbitals are taken
into account. In this model, the inter-V hopping is given as
$t_0+\delta t$ on the thick purple bond, while that
on other bonds is $t_0-\delta t$, where $t_0 <0$ denotes the
original hopping, and the BO parameter is $\delta t>0$.
Interestingly, the magnitude of the inter-V hopping
$|t_0+\delta t|$ decreases as the V-V bond length shortens.
This counterintuitive behavior arises from the
competition between direct V-V hopping $t_{\rm direct}$ and indirect V-Sb-V
hopping $t_{\rm indirect}$.
The inter-V hopping is divided into $t_{\rm direct}$ and $t_{\rm indirect}$.  In the original hopping, we have $t_0=t_{\rm direct} +t_{\rm indirect}<0$. 
In this case, we confirm the relations $t_{\rm direct}>0$, $t_{\rm indirect}<0$, and $|t_{\rm direct}|<|t_{\rm indirect}|$. 
In general, $|t_{\rm direct}|$ and $|t_{\rm indirect}|$ exhibit different dependences on the V–V bond length.
In the kagome metals, $|t_0+\delta t|$ decreases $(\delta t>0)$ when the V-V bond
length shortens, due to the competition between $t_{\rm direct}$ and
$t_{\rm indirect}$.
Similar results are known for
Fe-based superconductors, where the direct Fe-Fe hopping $t_{\rm direct}>0$ competes with
the indirect Fe-As-Fe hopping $t_{\rm indirect}<0$ \cite{Kuroki-dir-ind}.
The competition between $t_{\rm direct}$ and
$t_{\rm indirect}$ depends on the Fe-As-Fe angle.

%

\section{Summary}

In this paper, we studied the 3D nature of the DW in kagome metals
based on the 3D multiorbital models for AV$_3$Sb$_5$ (A=Cs, Rb, K),
by using the DW equation method. 
The TrH or SoD shape commensurate BO is robustly induced in each V-Sb plane 
by the PMI mechanism
that is described by the AL-VCs. $T_{\rm BO}\sim 100$K
is obtained in the moderate electron correlation.
Based on the analysis of the DW equation, by taking into
 account a finite third-order GL term, (i) $2\times 2\times 2$
s-BO can be realized via a first-order transition below
 $T_{\mathrm{BO}}$. Here, the in-plane BO pattern  (TrH or SoD) is
 determined by the sign of the third-order GL term, with hole doping
 tending to favor the TrH state. On the other hand, if the third-order
 GL term is very small, (ii) alternating v-BO may
 instead be realized via a second-order transition.
The rich variety of BOs observed in experiments can be understood by the
combination of the DW equation analysis and the 3rd-order GL free energy.
It was confirmed that the PMI mechanism is the essential origin of the 3D CDW of kagome metals.

\acknowledgements
This work was supported by JSPS
KAKENHI Grant Numbers JP25H01246, JP25H01248,
JP24K00568, JP24K06938, JP23K03299, JP22K14003.
\appendix

\section{3D Fermi Surface and Band Structure}

\begin{figure}[htb]
\includegraphics[width=.99\linewidth]{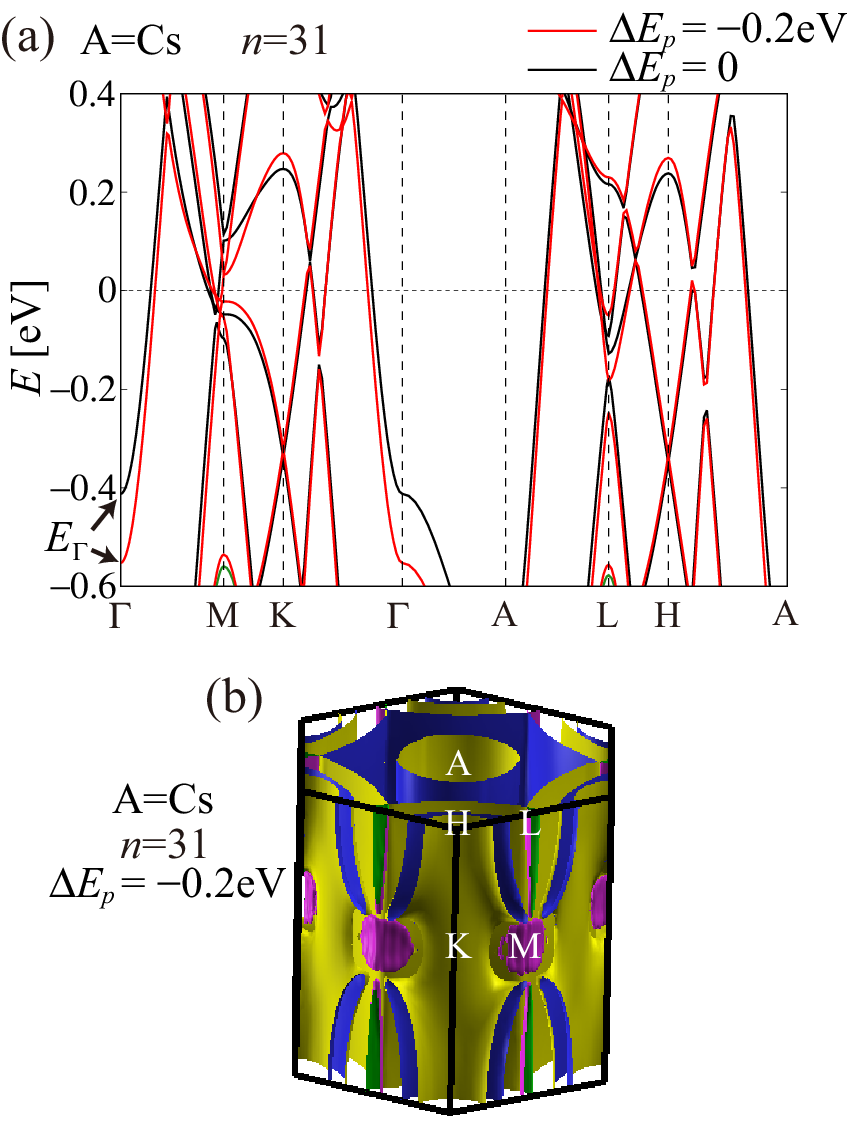}
\caption{
(a) Band structure of A=Cs for $\Delta E_P=0$ (black line) and
 that for $\Delta E_P=-0.2$ (red line).
(b) Three-dimensional FSs of A=Cs for $\Delta E_P=-0.2$.
}
\label{fig:figS2}
\end{figure}

\begin{figure}[htb]
\includegraphics[width=.99\linewidth]{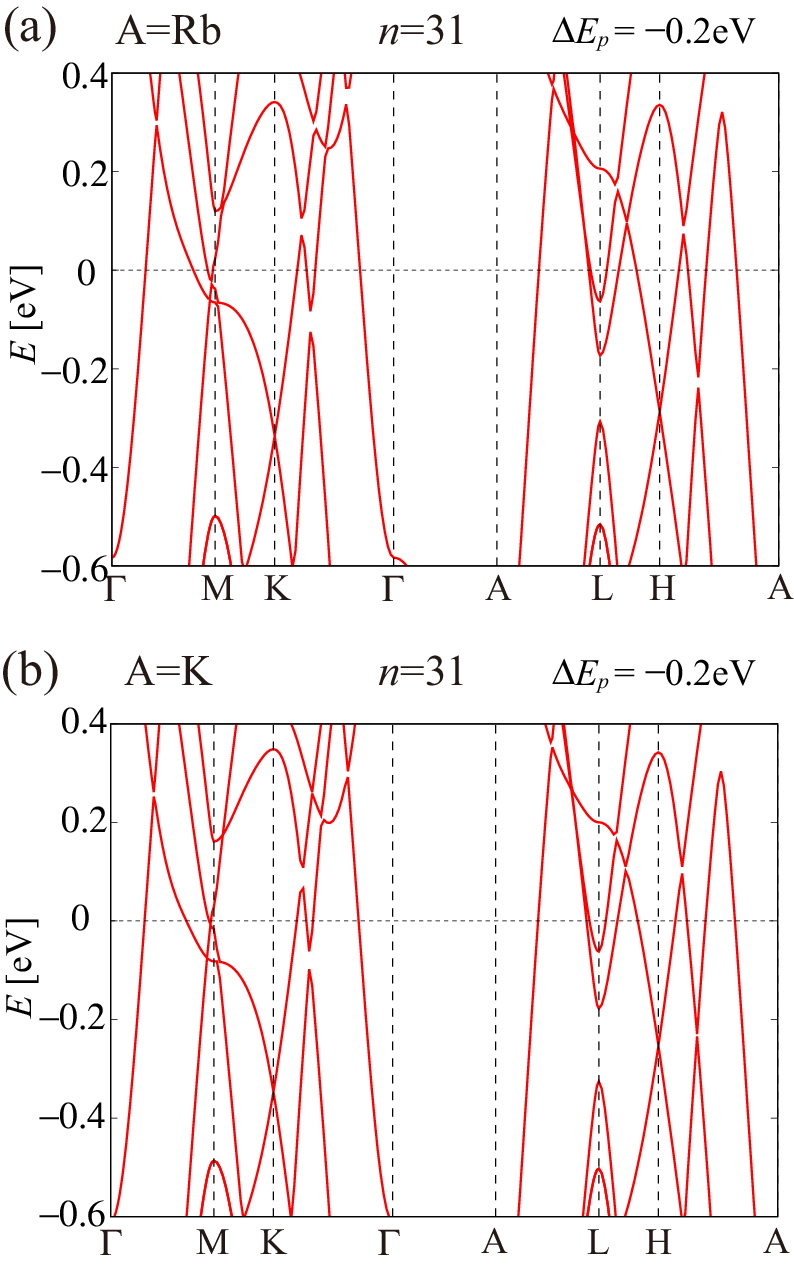}
\caption{
(a) Band structure of A=Rb for $\Delta E_P=-0.2$ and 
(b) that of A=K.
}
\label{fig:figS8}
\end{figure}

\begin{figure}[htb]
\includegraphics[width=.99\linewidth]{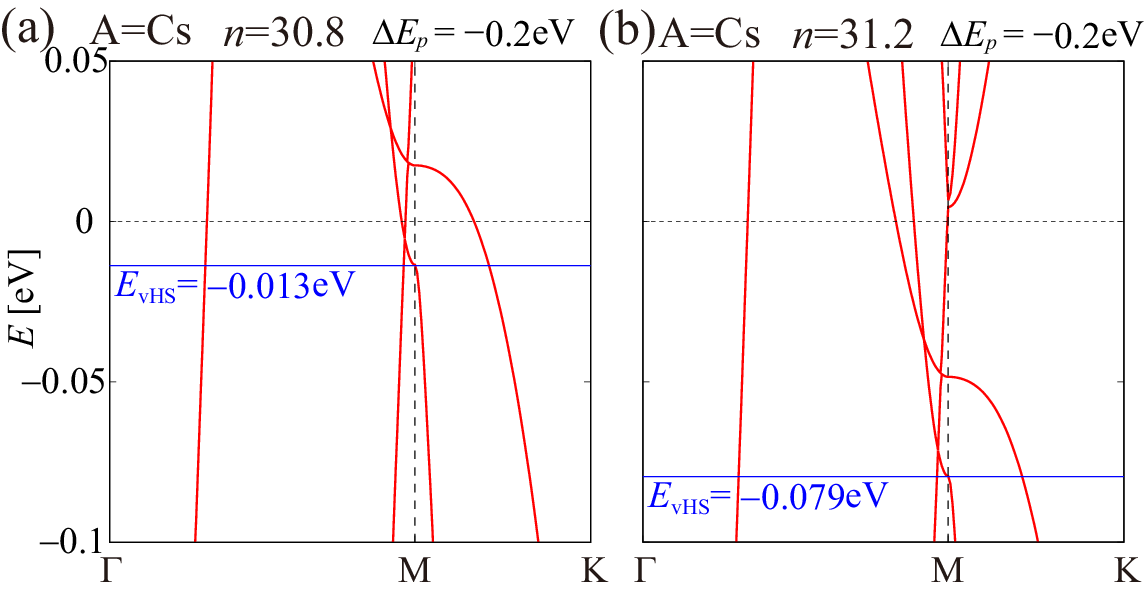}
\caption{
Band structures of A=Cs for $\Delta E_P=-0.2$ at (a) $n=30.8$ and (b) $n=31.2$.
Energy at the $b_{3g}$-orbital vHS $E_{\rm vHS}=-0.013(0.079)$ at $n=30.8(31.2)$.
}
\label{fig:fig15}
\end{figure}

We derive the 30 orbital tight-binding model with 
15 V $d$-orbitals and 15 Sb $p$-orbitals by using the WIEN2k \cite{WIEN2k} and Wannier90
\cite{Wannier90} software.
Figure \ref{fig:figS2} (a) shows the band structures for $\Delta E_p=-0.2,0$.
The band structure around the $\Gamma$ point is composed of Sb $p$-orbitals.
By introducing $\Delta E_p=-0.2$, energy at the $\Gamma$ point $E_\Gamma$
decreases from $-0.4$eV 
to $-0.55$eV consistently with ARPES measurements
\cite{exp8,ARPES-VHS}. Figure \ref{fig:figS2} (b) shows the
3D FS for $\Delta E_p=-0.2$ at $n=31$.
Figures \ref{fig:figS8} (a) and (b) show the band structures of A=Rb and
A=K for $\Delta E_p=-0.2$, respectively. In these cases, $E_\Gamma\sim -0.6$ is also consistent with ARPES measurements \cite{exp8}.
{Figures \ref{fig:fig15}(a) and (b) show the band structures of A=Cs for $\Delta E_p=-0.2$ at $n=30.8$ and $n=31.2$, respectively.
The energy at the $b_{3g}$-orbital vHS $E_{\rm vHS}$ is obtained as $E_{\rm vHS}=-0.013(0.079)$ at $n=30.8(31.2)$.}

\section{3rd-Order GL Coefficient}

\begin{figure}[htb]
\includegraphics[width=.99\linewidth]{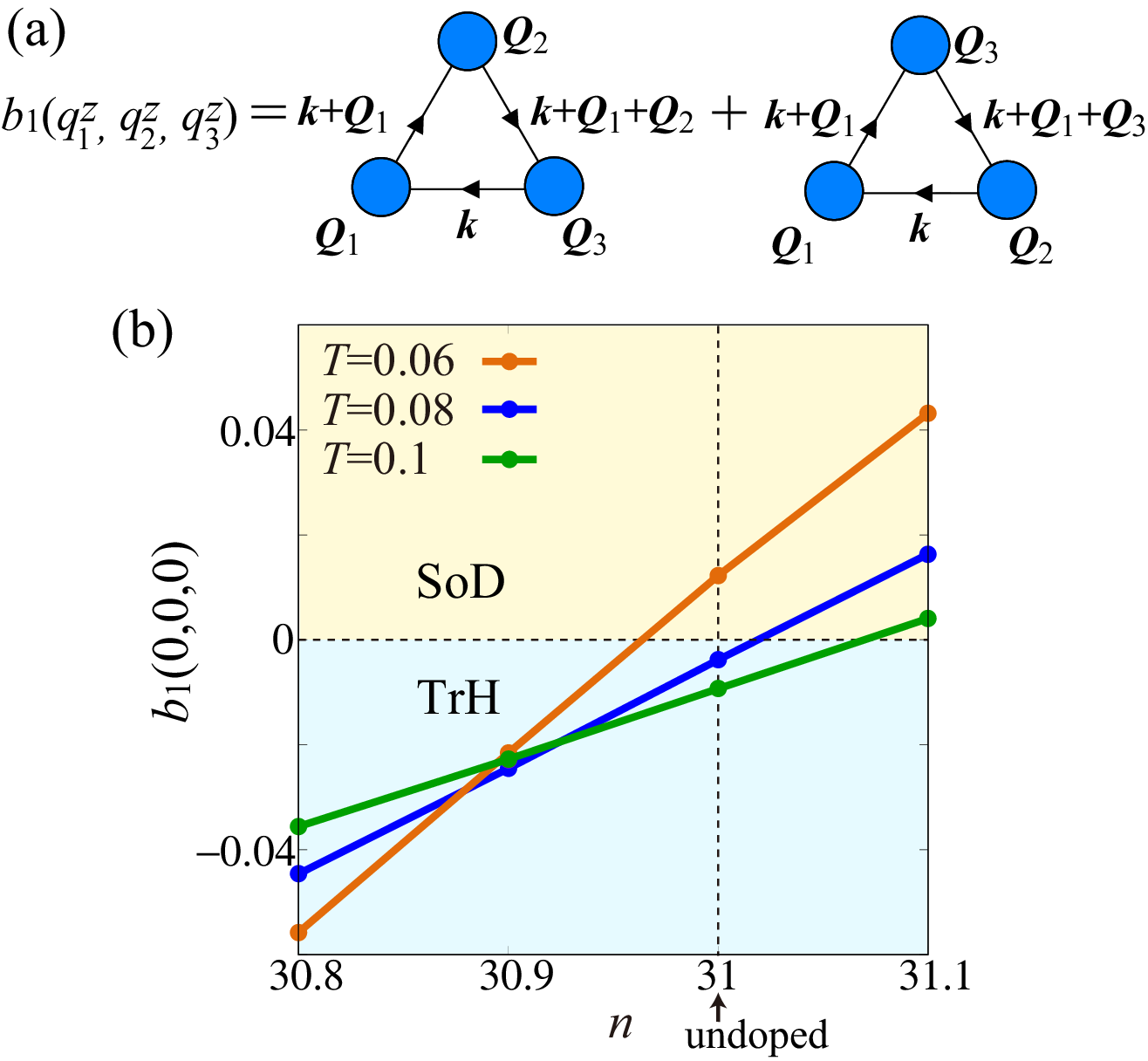}
\caption{
(a) Diagrammatic expression of the 3rd-order GL coefficient $b_1(q_1^z,q_2^z,q_3^z)$.
(b) $b_1(0,0,0)$ as functions of $n$ at $T=0.06$, $0.08$, and $0.1$ for A=Cs.
}
\label{fig:fig6}
\end{figure}

To understand the 3D $3Q$ BO state,
we also study the 3rd-order GL free energy $F^{(3)}=
b_1(q_1^z,q_2^z,q_3^z)\phi_1\phi_2\phi_3$ in the 3D kagome metal models.
${\bm\phi}=(\phi_1,\phi_2,\phi_3)$ are the order parameters, and the potential of the BO state at $\Q=\Q_m$ is given as $\phi_m{\hat g}_{\Q_m}(k)$.
Here, we fix the 2D BO wavevectors $\q_m$
while their $q_z$ components $q_m^z$ are variables,
under the constraint $q_1^z+q_2^z+q_3^z=0$ (modulo $2\pi$).
Here, we take account of commensurate $q_m^z=-q_m^z(=0,\pi)$
BO wavevectors. Diagrammatic expression of $b_1(q_1^z,q_2^z,q_3^z)$ is
given in Fig. \ref{fig:fig6} (a), whose analytical expression is
introduced in Ref. \cite{Tazai-kagome2}.

Figure \ref{fig:fig6} (b) shows
the numerical results of $b_1(0,0,0)$ at $T=0.06$, $0.08$, and $0.1$ for A=Cs. The obtained negative (positive) $b_1$ favors the TrH (SoD) BO.
We confirmed that $b_1(\pi,\pi,0)$ is almost the same as $b_1(0,0,0)$
since $q^z$ dependence of $b_1$ is very small.
Thus, $2\times 2\times 2$ TrH (SoD) s-BO competes with $2\times 2\times
1$ TrH (SoD) BO. 
From the $n$ dependence of $b_1$, the TrH BO tends to be favored in the hole-doped regime, while the SoD BO tends to be favored in the electron-doped regime.

We also confirmed that the $q^z$ dependence of the 4th-order GL coefficients is
smaller than that of the 3rd-order ones. Thus, the 3D structure of CDW
would be determined by the $q^z$-dependence of $\lambda_\Q$ and $b_1$.

\begin{thebibliography}{99}

\bibitem{kagome-exp1}
B. R. Ortiz, L. C. Gomes, J. R. Morey, M. Winiarski, M. Bordelon, J. S. Mangum, I. W. H. Oswald, J. A. Rodriguez-Rivera, J. R. Neilson, S. D. Wilson, E. Ertekin, T. M. McQueen, and E. S. Toberer,
{\it New kagome prototype materials: discovery of ${{KV}}_{3}{{Sb}}_{5},{{RbV}}_{3}{{Sb}}_{5}$, and ${{CsV}}_{3}{{Sb}}_{5}$},
{Phys. Rev. Materials} {\bf 3}, 094407 (2019).

\bibitem{kagome-exp2}
B. R. Ortiz, S. M. L. Teicher, Y. Hu, J. L. Zuo, P. M. Sarte, E. C. Schueller, A. M. M. Abeykoon, M. J. Krogstad, S. Rosenkranz, R. Osborn, R. Seshadri, L. Balents, J. He, and S. D. Wilson,
{\it ${Cs}{{V}}_{3}{{Sb}}_{5}$: A ${\mathbb{Z}}_{2}$ Topological Kagome Metal with a Superconducting Ground State},
{Phys. Rev. Lett.} {\bf 125}, 247002 (2020).

\bibitem{kagome-P-Tc1}
F. H. Yu, D. H. Ma, W. Z. Zhuo, S. Q. Liu, X. K. Wen, B. Lei, J. J. Ying, and X. H. Chen,
{\it Unusual competition of superconductivity and charge-density-wave state in a compressed topological kagome metal},
{Nat. Commun.} {\bf 12}, 3645 (2021).



\bibitem{STM1}
Y.-X. Jiang, J.-X. Yin, M. M. Denner, N. Shumiya, B. R. Ortiz, G. Xu, Z. Guguchia, J. He, M. S. Hossain, X. Liu, J. Ruff, L. Kautzsch, S. S. Zhang, G. Chang, I. Belopolski, Q. Zhang, T. A. Cochran, D. Multer, M. Litskevich, Z.-J. Cheng, X. P. Yang, Z. Wang, R. Thomale, T. Neupert, S. D. Wilson, and M. Z. Hasan,
{\it Unconventional chiral charge order in kagome superconductor KV$_{3}$Sb$_{5}$},
{Nat. Mater.} {\bf 20}, 1353 (2021).

\bibitem{STM2}
H. Li, H. Zhao, B. R. Ortiz, T. Park, M. Ye, L. Balents, Z. Wang, S. D. Wilson, and I. Zeljkovic,
{\it Rotation symmetry breaking in the normal state of a kagome superconductor KV$_{3}$Sb$_{5}$},
{Nat. Phys.} {\bf 18}, 265 (2022).

\bibitem{Thomale2013}
M. L. Kiesel, C. Platt, and R. Thomale,
{\it Unconventional Fermi Surface Instabilities in the Kagome Hubbard Model},
{Phys. Rev. Lett.} {\bf 110}, 126405 (2013).

\bibitem{SMFRG}
W.-S. Wang, Z.-Z. Li, Y.-Y. Xiang, and Q.-H. Wang,
{\it Competing electronic orders on kagome lattices at van Hove filling},
{Phys. Rev. B} {\bf 87}, 115135 (2013).

\bibitem{phonon-kagome}
H. Tan, Y. Liu, Z. Wang, and B. Yan,
{\it Charge Density Waves and Electronic Properties of Superconducting Kagome Metals},
{Phys. Rev. Lett.} {\bf 127}, 046401 (2021).

\bibitem{Thomale2021}
X. Wu, T. Schwemmer, T. M\"uller, A. Consiglio, G. Sangiovanni, D. Di Sante, Y. Iqbal, W. Hanke, A. P. Schnyder, M. M. Denner, M. H. Fischer, T. Neupert, and R. Thomale,
{\it Nature of Unconventional Pairing in the Kagome Superconductors $A{{V}}_{3}{{Sb}}_{5}$ ($A={K},{Rb},{Cs}$)},
{Phys. Rev. Lett.} {\bf 127}, 177001 (2021).

\bibitem{Neupert2021}
M. M. Denner, R. Thomale, and T. Neupert,
{\it Analysis of Charge Order in the Kagome Metal $A{{V}}_{3}{{Sb}}_{5}$ ($A={K},{Rb},{Cs}$)},
{Phys. Rev. Lett.} {\bf 127}, 217601 (2021).

\bibitem{Balents2021}
T. Park, M. Ye, and L. Balents,
{\it Electronic instabilities of kagome metals: Saddle points and Landau theory},
{Phys. Rev. B} {\bf 104}, 035142 (2021).

\bibitem{Nandkishore}
Y.-P. Lin and R. M. Nandkishore,
{\it Complex charge density waves at Van Hove singularity on hexagonal lattices: Haldane-model phase diagram and potential realization in the kagome metals AV$_{3}$Sb$_{5}$ (A = K, Rb, Cs)},
Phys. Rev. B {\bf 104}, 045122 (2021).

\bibitem{Tazai-kagome}
R. Tazai, Y. Yamakawa, S. Onari, and H. Kontani,
{\it Mechanism of exotic density-wave and beyond-Migdal unconventional superconductivity in kagome metal AV$_{3}$Sb$_{5}$ (A = K, Rb, Cs)},
{Sci. Adv.} {\bf 8}, eabl4108 (2022).

\bibitem{Tazai-kagome2}
R. Tazai, Y. Yamakawa, and H. Kontani,
{\it Charge-loop current order and Z3 nematicity mediated by bond-order fluctuations in kagome metals},
Nat. Commun. {\bf 14}, 7845 (2023).

\bibitem{Roppongi}
M. Roppongi, K. Ishihara, Y. Tanaka, K. Ogawa, K. Okada, S. Liu, K. Mukasa, Y. Mizukami, Y. Uwatoko, R. Grasset, M. Konczykowski, B. R. Ortiz, S. D. Wilson, K. Hashimoto, and T. Shibauchi,
{\it Bulk evidence of anisotropic s-wave pairing with no sign change in the kagome superconductor CsV$_3$Sb$_5$},
Nat. Commun. {\bf 14}, 667 (2023).

\bibitem{SC2}
W. Zhang, X. Liu, L. Wang, C. Wai T., Z. Wang, S. T. Lam, W. Wang, J. Xie, X. Zhou, Y. Zhao, S. Wang, J. Tallon, K. T. Lai, and S. K. Goh,
{\it Nodeless superconductivity in kagome metal CsV$_3$Sb$_5$ with and without time reversal symmetry breaking},
Nano Lett., {\bf 23}, 872 (2023).


\bibitem{Kohn}
H. Li, T. T. Zhang, T. Yilmaz, Y. Y. Pai, C. E. Marvinney, A. Said, Q. W. Yin, C. S. Gong, Z. J. Tu, E. Vescovo, C. S. Nelson, R. G. Moore, S. Murakami, H. C. Lei, H. N. Lee, B. J. Lawrie, and H. Miao,
{\it Observation of Unconventional Charge Density Wave without Acoustic Phonon Anomaly in Kagome Superconductors ${A{V}}_{3}{{Sb}}_{5}$ ($A={Rb}$, Cs)},
{Phys. Rev. X} {\bf 11}, 031050 (2021).

\bibitem{Kontani-AdvPhys}
H. Kontani, R. Tazai, Y. Yamakawa, and S. Onari,
{\it Unconventional density waves and superconductivities in Fe-based superconductors and other strongly correlated electron systems},
Adv. Phys. {\bf 70}, 355 (2021).

	\bibitem{muSR2-K}
	C. Mielke, D. Das, J.-X. Yin, H. Liu, R. Gupta, Y.-X. Jiang, M. Medarde, X. Wu, H. C. Lei, J. Chang, P. Dai, Q. Si, H. Miao, R. Thomale, T. Neupert, Y. Shi, R. Khasanov, M. Z. Hasan, H. Luetkens, and Z. Guguchia,
	{\it Time-reversal symmetry-breaking charge order in a kagome superconductor},
	Nature {\bf 602}, 245 (2022).
	
	\bibitem{muSR4-Cs}
	R. Khasanov, D. Das, R. Gupta, C. Mielke, M. Elender, Q. Yin, Z. Tu, C. Gong, H. Lei, E. T. Ritz, R. M. Fernandes, T. Birol, Z. Guguchia, and H. Luetkens,
	{\it Time-reversal symmetry broken by charge order in ${\mathrm{CsV}}_{3}{\mathrm{Sb}}_{5}$},
	Phys. Rev. Research {\bf 4}, 023244 (2022).
	
	\bibitem{muSR5-Rb}
	Z. Guguchia, C. Mielke, D. Das, R. Gupta, J.-X. Yin, H. Liu, Q. Yin, M. H. Christensen, Z. Tu, C. Gong, N. Shumiya, M. S. Hossain, T. Gamsakhurdashvili, M. Elender, P. Dai, A. Amato, Y. Shi, H. C. Lei, R. M. Fernandes, M. Z. Hasan, H. Luetkens, and R. Khasanov,
	{\it Tunable unconventional kagome superconductivity in charge ordered RbV$_3$Sb$_5$ and KV$_3$Sb$_5$},
	Nat. Commun. {\bf 14}, 153 (2023).

	\bibitem{AHE1}
S.-Y. Yang, Y. Wang, B. R. Ortiz, D. Liu, J. Gayles, E. Derunova, R. Gonzalez-Hernandez, L. $\check{S}$mejkal, Y. Chen, S. S. P. Parkin, S. D. Wilson, E. S. Toberer, T. McQueen, and M. N. Ali,
{\it Giant, unconventional anomalous Hall effect in the metallic frustrated magnet candidate, KV$_{3}$Sb$_{5}$},
{Sci. Adv.} {\bf 6}, eabb6003 (2020).

\bibitem{AHE2}
F. H. Yu, T. Wu, Z. Y. Wang, B. Lei, W. Z. Zhuo, J. J. Ying, and X. H. Chen,
{\it Concurrence of anomalous Hall effect and charge density wave in a superconducting topological kagome metal},
{Phys. Rev. B} {\bf 104}, L041103 (2021).

\bibitem{eMChA}
C. Guo, C. Putzke, S. Konyzheva, X. Huang, M. Gutierrez-Amigo, I. Errea, D. Chen, M. G. Vergniory, C. Felser, M. H. Fischer, T. Neupert, and P. J. W. Moll,
{\it Switchable chiral transport in charge-ordered Kagome metal CsV$_3$Sb$_5$},
Nature {\bf 611}, 461 (2022).

\bibitem{Moll-hz}
C. Guo, G. Wagner, C. Putzke, D. Chen, K. Wang, L. Zhang, M. G. Amigo, I. Errea, M. G. Vergniory, C. Felser, M. H. Fischer, T. Neupert, and P. J. W. Moll,
{\it Correlated order at the tipping point in the kagome metal CsV$_3$Sb$_5$},
Nat. Phys. {\bf 20}, 579 (2024).

\bibitem{Asaba-torque}
	T. Asaba, A. Onishi, Y. Kageyama, T. Kiyosue, K. Ohtsuka, S. Suetsugu, Y. Kohsaka, T. Gaggl, Y. Kasahara, H. Murayama, K. Hashimoto, R. Tazai, H. Kontani, B. R. Ortiz, S. D. Wilson, Q. Li, H. -H. Wen, T. Shibauchi, Y. Matsuda
	{\it Evidence for an odd-parity nematic phase above the charge density wave transition in kagome metal CsV$_3$Sb$_5$}.
Nat. Phys. {\bf 20}, 40 (2024).

	
\bibitem{Fernandes-CLC}
M. H. Christensen, T. Biro, B. M. Andersen, and R. M. Fernandes,
{\it Loop currents in AV$_3$Sb$_5$ kagome metals: Multipolar and toroidal magnetic orders},
Phys. Rev. B {\bf 106}, 144504 (2022).

\bibitem{Nat-g-ology}
H. D. Scammell, J. Ingham, T. Li, and O. P. Sushkov,
{\it Chiral excitonic order from twofold van Hove singularities in kagome metals},
Nat. Commun. {\bf 14}, 605 (2023).

\bibitem{Onari-FeSe}
S. Onari, Y. Yamakawa, and H. Kontani,
{\it Sign-Reversing Orbital Polarization in the Nematic Phase of FeSe due to the ${C}_{2}$ Symmetry Breaking in the Self-Energy},
{Phys. Rev. Lett.} {\bf 116}, 227001 (2016).

\bibitem{Onari-AFN}
S. Onari and H. Kontani,
{\it Hidden antiferronematic order in Fe-based superconductor ${Ba}{{Fe}}_{2}{{As}}_{2}$ and NaFeAs above ${T}_{S}$},
{Phys. Rev. Research} {\bf 2}, 042005(R) (2020).

\bibitem{Onari-Ni}
S. Onari and H. Kontani,
{\it Three-dimensional bond-order instability in infinite-layer nickelates due to nonlocal quantum interference},
{Phys. Rev. B} {\bf 108}, L241119 (2023).

\bibitem{exp13}
H. Luo, Q. Gao, H. Liu, Y. Gu, D. Wu, C. Yi, J. Jia, S. Wu, X. Luo,
Y. Xu, L. Zhao, Q. Wang, H. Mao, G. Liu, Z. Zhu, Y. Shi, K. Jiang,
J. Hu, Z. Xu, and X. J. Zhou,
{\it Electronic nature of charge density wave and electron-phonon coupling in kagome superconductor KV$_3$Sb$_5$},
Nat. Commun. {\bf 13}, 273 (2022).
\bibitem{exp15}
Z. Wang, Y.-X. Jiang, J.-X. Yin, Y. Li, G.-Y. Wang,
H.-L. Huang, S. Shao, J. Liu, P. Zhu, N. Shumiya, M. S. Hossain, H. Liu,
Y. Shi, J. Duan, X. Li, G. Chang, P. Dai, Z. Ye, G. Xu, Y. Wang,
H. Zheng, J. Jia, M. Z. Hasan, and Y. Yao,
{\it Electronic nature of chiral charge order in the kagome superconductor CsV$_3$Sb$_5$},
Phys. Rev. B {\bf 104}, 075148 (2021).
\bibitem{exp16}
L. Nie, K. Sun, W. Ma, D. Song, L. Zheng, Z. Liang, P. Wu, F. Yu, J. Li,
M. Shan, D. Zhao, S. Li, B. Kang, Z. Wu, Y. Zhou, K. Liu, Z. Xiang,
J. Ying, Z. Wang, T. Wu, and X. Chen,
 {\it Charge-density-wave-driven electronic nematicity in a kagome superconductor},
Nature {\bf 604}, 59 (2022).
\bibitem{exp17}
T. Kato, Y. Li, T. Kawakami, M. Liu, K. Nakayama, Z. Wang, A. Moriya,
K. Tanaka, T. Takahashi, Y. Yao, and T. Sato,
{\it Three-dimensional energy gap and origin of charge-density wave in kagome superconductor KV$_3$Sb$_5$},
Commun. Mater. {\bf 3}, 30 (2022).
\bibitem{exp18}
J. Luo, Z. Zhao, Y. Z. Zhou, J. Yang, A. F. Fang, H. T. Yang, H. J. Gao,
R. Zhou, and G.-Q. Zheng,
{\it Possible star-of-David pattern charge density wave with additional modulation in the kagome superconductor CsV$_3$Sb$_5$},
npj Quantum Mater. {\bf 7}, 30 (2022).
\bibitem{exp19}
Z. Liu, N. Zhao, Q. Yin, C. Gong, Z. Tu, M. Li, W. Song, Z. Liu, D. Shen, Y. Huang, K. Liu, H. Lei, and S. Wang,
{\it Charge-Density-Wave-Induced Bands Renormalization and Energy Gaps in a Kagome Superconductor RbV$_3$Sb$_5$},
Phys. Rev. X {\bf 11}, 041010 (2021).
\bibitem{exp20}
S. Cho, H. Ma, W. Xia, Y. Yang, Z. Liu, Z. Huang, Z. Jiang, X. Lu, J. Liu, Z. Liu, J. Li, J. Wang, Y. Liu, J. Jia, Y. Guo, J. Liu, and D. Shen,
{\it Emergence of New van Hove Singularities in the Charge Density Wave State of a Topological Kagome Metal RbV$_3$Sb$_5$},
Phys. Rev. Lett. {\bf 127}, 236401 (2021).
\bibitem{exp21}
Y. Hu, X. Wu, B. R. Ortiz, X. Han, N. C. Plumb, S. D. Wilson, A. P. Schnyder, and M. Shi,
{\it Coexistence of trihexagonal and star-of-David pattern in the charge density wave of the kagome superconductor AV$_3$Sb$_5$},
Phys. Rev. B {\bf 106}, L241106 (2022).

\bibitem{exp12}
Q. Stahl, D. Chen, T. Ritschel, C. Shekhar, E. Sadrollahi, M. C. Rahn, O. Ivashko, M. v. Zimmermann, C. Felser, and J. Geck,
{\it Temperature-driven reorganization of electronic order in CsV$_3$Sb$_5$},
Phys. Rev. B {\bf 105}, 195136 (2022).


\bibitem{exp5}
Z. Liang, X. Hou, F. Zhang, W. Ma, P. Wu, Z. Zhang, F. Yu, J.-J. Ying, K. Jiang, L. Shan, Z. Wang, and X.-H. Chen,
{\it Three-Dimensional Charge Density Wave and Surface-Dependent
Vortex-Core States in a Kagome Superconductor CsV$_3$Sb$_5$},
Phys. Rev. X {\bf 11}, 031026 (2021).
\bibitem{exp2}
Y. Wang, T. Wu, Z. Li, K. Jiang, and J. Hu, 
{\it Structure of the kagome superconductor CsV$_3$Sb$_5$ in the charge density wave state}, 
Phys. Rev. B {\bf 107}, 184106 (2023).
\bibitem{exp3}
J. Frassineti, P. Bonf\'{a}, G. Allodi, E. Garcia, R. Cong, B. R. Ortiz, S. D. Wilson, R. D. Renzi, V. F. Mitrovi\'{c}, and S. Sanna, 
{\it Microscopic nature of the charge-density wave in the kagome superconductor RbV$_3$Sb$_5$},
Phys. Rev. Research {\bf 5}, L012017 (2023).
\bibitem{exp4}
C. Mu, Q. Yin, Z. Tu, C. Gong, P. Zheng, H. Lei, Z. Li, and J. Luo,
{\it Tri-Hexagonal charge order in kagome metal CsV$_3$Sb$_5$ revealed by $^{121}$Sb NQR},
Chin. Phys. B {\bf 31}, 017105 (2022).
\bibitem{exp6}
D. W. Song, L. X. Zheng, F. H. Yu, J. Li, L. P. Nie, M. Shan, D. Zhao, S. J. Li, B. L. Kang, Z. M. Wu, Y. B. Zhou, K. L. Sun, K. Liu, X. G. Luo, Z. Y. Wang, J. J. Ying, X. G. Wan,T. Wu, and X. H. Chen,
{\it Orbital ordering and fluctuations in a kagome superconductor
CsV$_3$Sb$_5$}, Sci. China. Phys. Mech. Astron. {\bf 65}, 247462 (2022).

\bibitem{exp1}
Z. Jiang, H. Ma, W. Xia, Z. Liu, Q. Xiao, Z. Liu, Y. Yang, J. Ding, Z. Huang, J. Liu, Y. Qiao, J. Liu, Y. Peng, S. Cho, Y. Guo, J. Liu, and D. Shen, 
{\it Observation of Electronic Nematicity Driven by the Three Dimensional Charge Density Wave in Kagome Lattice KV$_3$Sb$_5$},
Nano Lett. {\bf 23}, 5625 (2023).


 \bibitem{exp8}
M. Kang, S. Fang, J. Yoo, B. R. Ortiz, Y. M. Oey, J. Choi, S. H. Ryu,
J. Kim, C. Jozwiak, A. Bostwick, E. Rotenberg, E. Kaxiras,
J. G. Checkelsky, S. D. Wilson, J.-H. Park, and R. Comin,
 {\it Charge order landscape and competition with superconductivity in kagome metals},
 Nat. Mater. {\bf 22}, 186 (2023).

 \bibitem{Raman}
F. Jin, W. Ren, M. Tan, M. Xie, B. Lu, Z. Zhang,
J. Ji , and Q. Zhang,
{\it $\pi$ Phase Interlayer Shift and Stacking Fault in the Kagome
Superconductor CsV$_3$Sb$_5$},
Phys. Rev. Lett. {\bf 132}, 066501 (2024).

 \bibitem{exp9}
H. Li, G. Fabbris, A. H. Said, J. P. Sun, Y.-X. Jiang, J.-X. Yin,
Y.-Y. Pai, S. Yoon, A. R. Lupini, C. S. Nelson, Q. W. Yin, C. S. Gong,
Z. J. Tu, H. C. Lei, J.-G. Cheng, M. Z. Hasan, Z. Wang, B. Yan,
R. Thomale, H. N. Lee, and H. Miao,
 {\it Discovery of conjoined charge density waves in the kagome superconductor CsV$_3$Sb$_5$},
Nat. Commun. {\bf 13}, 6348 (2022). 

\bibitem{XRD}
L. Kautzsch, B. R. Ortiz, K. Mallayya, J. Plumb, G. Pokharel, J. P. C. Ruff, Z. Islam, E.-A. Kim, R. Seshadri, S. D. Wilson,
{\it Structural evolution of the kagome superconductors AV$_3$Sb$_5$ (A = K, Rb, and Cs) through charge density wave order},
Phys. Rev. Materials {\bf 7}, 024806 (2023).

\bibitem{exp22}
Q. Xiao, Y. Lin, Q. Li, X. Zheng, S. Francoual, C. Plueckthun, W. Xia,
Q. Qiu, S. Zhang, Y. Guo, J. Feng, and Y. Peng,
{\it Coexistence of multiple stacking charge density waves in kagome
superconductor CsV$_3$Sb$_5$},
Phys. Rev. Research {\bf 5}, L012032 (2023).

\bibitem{exp23}
J. Plumb, A. C. Salinas, K. Mallayya, E. Kisiel,
F. B. Carneiro, R. Gomez, G. Pokharel, E.-A. Kim,
S. Sarker, Z. Islam, S. Daly, S. D. Wilson,
{\it Phase-Separated Charge Order and Twinning Across Length Scales in
CsV$_3$Sb$_5$},
Phys. Rev. Materials {\bf 8}, 093601 (2024).

\bibitem{exp10}
B. R. Ortiz, S. M. L. Teicher, L. Kautzsch, P. M. Sarte, N. Ratcliff,
J. Harter, J. P. C. Ruff, R. Seshadri, and S. D. Wilson,
{\it Fermi Surface Mapping and the Nature of Charge-Density-Wave Order in the Kagome Superconductor CsV$_3$Sb$_5$},
Phys. Rev. X {\bf 11}, 041030 (2021).
\bibitem{exp11}
S. Wu, B. R. Ortiz, H. Tan, S. D. Wilson, B. Yan, T. Birol, and G. Blumberg,
{\it Charge density wave order in the kagome metal AV$_3$Sb$_5$
(A=Cs,Rb,K)},
Phys. Rev. B {\bf 105}, 155106 (2022).


\bibitem{NMR-SoD}
X. Y. Feng, Z. Zhao, J. Luo, J. Yang, A. F. Fang, H. T. Yang, H. J. Gao, R. Zhou,and G.-Q. Zheng,
{\it Commensurate-to-incommensurate transition of charge-
density-wave order and a possible quantum critical point
in pressurized kagome metal CsV$_3$Sb$_5$},
npj Quantum Mater. {\bf 8}, 23 (2023).

\bibitem{theo2}
Andrzej Ptok, Aksel Kobialka, Malgorzata Sternik, Jan La\.{z}ewski, Pawel T. Jochym, Andrzej M. Ole\'{s}, and Przemyslaw Piekarz,
{\it Dynamical study of the origin of the charge density wave in AV$_3$Sb$_5$ (A=K, Rb, Cs) compounds},
Phys. Rev. B {\bf 105}, 235134 (2022).
\bibitem{theo3}
C. Wang, S. Liu, H. Jeon, Y. Jia, and J.-H. Cho,
{\it Charge density wave and superconductivity in the kagome metal CsV$_3$Sb$_5$ around a pressure-induced quantum critical point},
Phys. Rev. Materials {\bf 6}, 094801 (2022).
\bibitem{theo4}
C. Wang, S. Liu, H. Jeon, and J.-H. Cho,
{\it Origin of charge density wave in the layered kagome metal CsV$_3$Sb$_5$},
Phys. Rev. B {\bf 105}, 045135(2022).


\bibitem{theo6}
H. Li, X. Liu, Y. B. Kim, and H.-Y. Kee,
{\it Origin of $\pi$-shifted three-dimensional charge density waves in the
kagome metal AV$_3$Sb$_5$ (A=Cs, Rb, K)},
Phys. Rev. B {\bf 108}, 075102 (2023).
\bibitem{theo7}
F. Grandi, A. Consiglio, M. A. Sentef, R. Thomale,
and D. M. Kennes,
{\it Theory of nematic charge orders in kagome metals},
Phys. Rev. B {\bf 107}, 155131 (2023).
\bibitem{theo9}
F. Ferrari, F. Becca, and R. Valenti,
{\it Charge density waves in kagome-lattice extended Hubbard models at the van Hove filling},
Phys. Rev. B {\bf 106}, L081107 (2022).

\bibitem{theo5}
M. H. Christensen, T. Birol, B. M. Andersen, and R. M. Fernandes,
{\it Theory of the charge density wave in AV$_3$Sb$_5$ kagome metals},
Phys. Rev. B {\bf 104}, 214513 (2021).

\bibitem{theo8}
E. T. Ritz, R. M. Fernandes, and T. Birol,
{\it Impact of Sb degrees of freedom on the charge density wave phase diagram of the kagome metal CsV$_3$Sb$_5$},
Phys. Rev. B {\bf 107}, 205131 (2023).


\bibitem{WIEN2k}
 P. Blaha, K. Schwarz, F. Tran, R. Laskowski, G. K. H.
 Madsen, and L. D. Marks,
 {\it WIEN 2K : An APW+lo Program
for Calculating the Properties of Solids}, J. Chem. Phys. {\bf 152},
074101 (2020).

\bibitem{Wannier90}
G. Pizzi {\it et al}.,
{\it Wannier90 as a community code: new features and
applications},
J. Phys. Cond. Matt. {\bf 32}, 165902 (2020).




\bibitem{ARPES-VHS}
Y. Hu, X. Wu, B. R. Ortiz, S. Ju, X. Han, J. Ma, N. C. Plumb, M. Radovic, R. Thomale, S. D. Wilson, A. P. Schnyder, and M. Shi,
{\it Rich nature of Van Hove singularities in Kagome superconductor CsV$_3$Sb$_5$},
Nat. Commun. {\bf 13}, 2220 (2022).


\bibitem{Kontani-PRL2010}
H. Kontani and S. Onari,
{\it Orbital-Fluctuation-Mediated Superconductivity in Iron Pnictides: Analysis of the Five-Orbital Hubbard-Holstein Model},
Phys. Rev. Lett. {\bf 104}, 157001 (2010).
\bibitem{Onari-PRL2012}
S. Onari and H. Kontani,
{\it Self-consistent Vertex Correction Analysis for Iron-based Superconductors:
Mechanism of Coulomb Interaction-Driven Orbital Fluctuations},
Phys. Rev. Lett. {\bf 109}, 137001 (2012).

\bibitem{Yamakawa-Cu}
Y. Yamakawa and H. Kontani,
{\it Spin-Fluctuation-Driven Nematic Charge-Density Wave in Cuprate Superconductors: Impact of Aslamazov-Larkin Vertex Corrections},
{Phys. Rev. Lett.} {\bf 114}, 257001 (2015).


\bibitem{Onari-TBG}
S. Onari and H. Kontani,
{\it SU(4) Valley + Spin Fluctuation Interference Mechanism for Nematic
Order in Magic-Angle Twisted Bilayer Graphene: The Impact of Vertex
Corrections},
Phys. Rev. Lett. {\bf 128}, 066401 (2022).

\bibitem{Fe-phonon}
T. Saito, S. Onari, and H. Kontani,
{\it Emergence of fully gapped $s++$-wave and nodal $d$-wave states mediated by orbital and spin fluctuations in a ten-orbital model of KFe$_2$Se$_2$},
Phys. Rev. B {\bf 83}, 140512(R) (2011).

\bibitem{Kuroki-dir-ind}
K. Suzuki, H. Usui, S. Iimura, Y. Sato, S. Matsuishi, H. Hosono, and
K. Kuroki,
Phys. Rev. Lett. {\bf 113}, 027002 (2014).


%
%
%
%
%
%
%
%
%
%
%
%
%
%
%
%
%
%
%
%
%
%
%

%
%





	

%
%
%
%
%
%
%
%
%
%
%
%
%
%
%
%
%
%
%
%
%



%
%
%
%


%
%
%
%
%
%
%
%
%
%
%
%













\end{thebibliography}
\end{document}